\documentclass[aps,prb,twocolumn,groupedaddress,floatfix,superscriptaddress]{revtex4-2}

\usepackage{epsfig}
\usepackage{graphicx}
\usepackage[normalem]{ulem}
\usepackage{mathtools}
\usepackage{amsmath,amssymb}
\usepackage{physics}
\usepackage{hyperref}
\hypersetup{colorlinks=true,citecolor={blue},linkcolor={blue},urlcolor={blue}}
\usepackage[dvipsnames,usenames]{xcolor}

\newcommand{\cqtadd}{Centre for Quantum Technologies, National
  University of Singapore, 3 Science Drive 2, 117543 Singapore}
\newcommand{\dpnusadd}{Department of Physics, National University of
  Singapore, 2 Science Drive 3, 117551 Singapore} 

\newcommand{\corrdisorder}{\xi_{\rm disorder}}

\begin{document}

\title{Disorder and the Robustness of Superconductivity on the Flat
  Band}

\author{Si Min Chan}
    \affiliation{\cqtadd}
    \affiliation{\dpnusadd} 
    \affiliation{Home Team Science and Technology Agency, 1 Stars Ave, 138507 Singapore}

\author{B. Gr\'emaud}
    \affiliation{Aix Marseille Univ, Universit\'e de Toulon, CNRS,
      CPT, Marseille, France} 

\author{G. George Batrouni}
\affiliation{Universit\'e C\^ote d'Azur, CNRS, Institut de Physique de
  Nice (INPHYNI), 06103 Nice, France}
\affiliation{New Cornerstone Science Laboratory, Department of
  Physics, School of Science, Westlake University, Hangzhou 310024,
  Zhejiang, China}

\begin{abstract}
We study the interplay between on-site disorder and fermion pairing on
the quasi one-dimensional flat band Creutz lattice. Both disorder and
flat bands localize particles, but an attractive interaction results
in pair formation and delocalization giving rise to
superconductivity. In this work, we examine the attractive Hubbard
model on the Creutz lattice to study the competition between these two
effects and elucidate the properties of the superconducting phase and
the localization quantum phase transition as the disorder strength is
increased. Our main result is that
flat band superconductivity is robust against disorder: The
critical disorder strength, $W_c$, required to localize the fermion
pairs and destroy superconductivity, is finite at any interaction
strength, $U$, and is proportional to the superconducting weight,
$D_s$, of the clean system. Using large scale density matrix
renormalization group computations, we show that this transition is of the BKT form. In addition, even at very small
interaction strength, the localization is not due to single fermion
localization but to pair localization. For completeness, we briefly study this disorder-induced localization with mean field theory and show that $W_c$ can be accurately determined by using an appropriate scaling function.

\end{abstract}

\maketitle 

\section{Introduction}

Interactions in flat band systems give rise to exotic strongly
correlated states and nontrivial order like ferromagnetism,
antiferromagnetism, superfluidity and
superconductivity~\cite{derzhko2015strongly, gulacsi2014interaction,
  zeng1990numerical,asakawa1994possibility,
  mielke1991ferromagnetic,mielke1991ferromagnetism,tasaki1992ferromagnetism,mielke1993ferromagnetism,scalapino1992superfluid,peotta2015superfluidity,chan2022pairing,
  chan2022designer,
  iskin2019origin,julku2016geometric,huhtinen2022revisiting}. Real
systems, however, can contain impurities or other types of disorder
like vacancies and dislocations. This raises an important question
about the robustness of these quantum phases in the presence of
disorder: Does any amount of disorder, no matter how weak, change the
nature of the quantum phase, or is a minimum disorder strength
required? In this work, we focus on the effect of local diagonal site
disorder on superconductivity in a quasi one-dimensional system with a
flat band in the ground state.

In a non-interacting $1D$ system with a dispersive band, the presence
of disorder leads to Anderson localization, which results from the
scattering of quantum particles and interference
effects~\cite{anderson1958absence, kramer1993localization}. In the
pure system, the wavefunctions are plane waves, but when disorder is
introduced, the particles are localized in pockets where the
wavefunction becomes tightly confined, resulting in complete
suppression of wave transport no matter how weak the disorder is. When
there are interactions between the particles, the presence of disorder
may lead to many-body
localization.\cite{pal2010,bardarson2012,luitz2015,imbrie2016,abanin2019}
If a certain degree of disorder or randomness is required to localize
the system, then the critical disorder strength, $W_c$, is
finite. Below the critical point, particles are still mobile and
transport, for example superconductivity, survives.

Localization also occurs on flat bands arising from a different
mechanism. Flat band systems have zero band width, resulting in a
vanishing group velocity in the non-interacting case. The
wavefunctions of non-interacting particles are localized due to
destructive interference and decay exponentially, or faster than
exponentially~\cite{mallick2021wannier}, or even strictly
compact~\cite{read2017compactly}. These states are fragile towards
nonzero perturbations, especially attractive interactions, which
destroy the macroscopic degeneracy and result in nontrivial phases.

The theoretical study of the interplay between disorder and
interaction is important in order to understand the robustness of
superconductivity and transport on flat bands. However, the behavior
near the critical point is still an open question, and there is little
in the literature about the interplay between disorder and interaction
in systems with spin and charge degrees of
freedom~\cite{khemani2017critical,panda2020can,alet2018many}. Additionally,
not much is known about the effect of weak disorder on spinful
fermionic many-body systems, especially on flat bands.

Existing studies on quasi one-dimensional dispersive bands considered
repulsive bosons which have a Berezinskii–Kosterlitz–Thouless (BKT)
transition between the localized to delocalized
phases~\cite{fisher1989boson,scalettar1991localization,michal2016finite,ristivojevic2012phase,hopjan2020many},
and infinitely repulsive (hard-core) bosons with a finite critical
disorder, below which there will be a superfluid
phase~\cite{carrasquilla2011bose, doggen2017weak}. Near neighbour
attractive interaction results in delocalization of spinless fermions,
even with disorder~\cite{schmitteckert1998fermi,
  schmitteckert1998anderson, lin2018many}, and spinful repulsive
fermions were studied with various hopping geometries and types of
disorder~\cite{mondaini2015many,prelovvsek2016absence,zakrzewski2018spin,
  sroda2019instability,bahovadinov2022many}.  It is important to point
out the difference in physics of a simple $1D$ chain and coupled
chains in quasi-$1D$, in particular for hard-core bosons and
fermions. With multiple coupled chains, the particles can exchange
among each other, in contrast with a simple single chain where the
motion is collective as particles cannot move around each other. This
results in a sustained superconducting state at finite disorder when there are multiple chains, a behavior different from localization at any
nonzero disorder on single
chains~\cite{giamarchi1988anderson,orignac1997effects,orignac1996effects,orignac1999,carrasquilla2011bose,crepin2011phase}. In
the present case, multiple chains are required to construct flat
bands, thus we consider the quasi-$1D$ disordered systems.

Several papers considered disorder on flat bands, but do not take into
account interaction~\cite{chalker2010anderson}, studied spinless
fermions~\cite{roy2020interplay}, with repulsive
interaction~\cite{orito2022deformation, orito2021interplay} or with mean field theory at weak attractive interaction \cite{lau2022universal}. Flat band
systems with weak attraction have finite wavefunction overlap between
the formed pairs, even though the single particle correlation length
is very short; in other words they are not truly hard-core bosons. In
a recent paper~\cite{Liang2023Disorder}, the authors analyzed
superconductivity on the Creutz ladder with on-site charge disorder,
$\mu_j (n_{j,\uparrow}+n_{j,\downarrow})$, and Zeeman-type spin
disorder, $h_j (n_{j,\uparrow}-n_{j,\downarrow})$, characterizing the
system with the Hubbard model at a relatively high attractive coupling
of $U=8t$ (twice the gap between the two bands of the Creutz lattice),
and using density matrix renormalization group (DMRG)
calculations. The authors found that with on-site disorder, $W_c=0$
for both flat (Creutz lattice) and dispersive (simple two-chain ladder
lattice) bands. They noted that the Luttinger parameter, $K$, for both
the Creutz and simple ladder lattices at $U=8t$, satisfies $K<K_c=3/2$
~\cite{orignac1996effects,orignac1997effects,orignac1999} and
therefore favors an instability to localization for any amount of
on-site disorder, $W_c=0$.  However, with Zeeman-type disorder there
is finite $W_c$ on the Creutz lattice while $W_c$ remains zero on the
simple ladder.

The Luttinger parameter can be understood as follows. In the pure
quasi-$1D$ system without disorder at zero temperature, interacting
bosons or fermions are described by the Luttinger liquid (LL) model
and can exist with quasi long-range
order~\cite{luttinger1963exactly,kuhner2000one}. Correlations decay as
a power law with $\langle b_j^\dagger b_{j+r} \rangle \sim r^{-1/2K}$
for bosons, and the pair Green's function $\langle
c_{j+r,\downarrow}^\alpha c_{j+r,\uparrow}^\alpha
c_{j,\uparrow}^{\alpha'\dagger}c_{j,\downarrow}^{\alpha'\dagger}\rangle
\sim r^{-\omega}=r^{-1/2K}$ for spin-1/2 fermions with attractive
interactions. The Luttinger parameter, $K$, typically depends on the
interaction strength and the details of the hopping parameter,
\textit{i.e.} the geometry. In the case of dispersive bands, bosonization
predicts (for weak interactions) critical values for the Luttinger
parameter, $K_c$, below which the system will localize for any
disorder strength and superfluidity is destroyed due to quantum phase
slips~\cite{orignac1996effects,orignac1997effects,kuhner2000one,doggen2017weak}.
On the simple two-chain ladder with a dispersive band,
$\omega=\tfrac{1}{2K}$ decreases with increasing attraction
strength~\cite{mondaini2018pairing}, but $K$ remains smaller than
$K_c=3/2$ for all values of $U$ investigated. Therefore, in this case,
localization is expected for any disorder
strength~\cite{Liang2023Disorder}, at least for small $U$ where
bosoniztion is
reliable. References~\cite{chan2022pairing,mondaini2018pairing,chan2022designer}
showed that for flat bands, $\omega=\tfrac{1}{2K}$ \textit{increases}
($K$ decreases) as $U$ increases (see below). This behavior is
qualitatively different from the dispersive band case and offers the
possibility that for small enough $U$, $K>K_c$ leading to a nonzero
critical value of the disorder, $W_c\neq 0$, needed to localize the
system. On the Creutz lattice at quarter (eighth) filling with
attractive on-site interaction, $K>K_c$ when $U\lesssim 10$
($U\lesssim
5$),~\cite{chan2022pairing,mondaini2018pairing,chan2022designer} and
$K<K_c$ when $U\gtrsim 10$ and $U\gtrsim 5$ respectively. Assuming
that at low $U$ values bosonization predictions apply to the flat
band, we conclude that for quarter (eighth) filling, the critical
value of disorder required to localize the system is nonzero, $W_c\neq
0$, for $U\lesssim 10$ ($U\lesssim 5$). In other words,
superconductivity (SC) is robust and persists even in the presence of
disorder. When $U$ is large enough, however, the Luttinger parameter
becomes less than the critical value, $K<K_c$, implying, perhaps, that
in this $U$ regime any amount of disorder will localize the system,
$W_c=0$. However, the situation is not so simple. Bosonization
predictions are reliable at low interaction strengths, and they depend
on the type of particle involved: spinful fermions, spinlesss
fermions, hard core bosons (HCB) etc. We note that for the attractive
Hubbard model at very strong interaction, the model is very well
described by hard core bosons with hopping parameter $t^2/U$ and an
equal repulsive near neighbor interaction (see below). Furthermore, at
such strong interaction, the flat band has disappeared and is replaced
by dispersive bands since in the effective HCB model all the hopping
terms have the same sign and there is no quantum
interference. Consequently, the critical Luttinger parameter at very
large $U$ is determined by the physics of hard core bosons on coupled
chains and has a value $K^{\rm HCB}_c=3/4$.\cite{crepin2011phase} We shall see
below that in our system $K > K^{\rm HCB}_c$ for any value of the
interaction and in particular at large $U$ leading, once again, to
robust SC persisting at finite disorder.

We, therefore, argue that at low and at large values of $U$ the system
superconductivity is robust and requires a finite value of the
disorder to be localized. What happens for intermediate values of $U$
where bosonization is not reliable? there are two possibilities. The
first is that as $U$ is increased, the system passes from a robust
phase with $W_c\neq 0$ to a fragile phase with $W_c=0$ and, as $U$ is
increased further, the system passes into an HCB phase where
superconductivity is again robust and $W_c\neq 0$.  The second
possibility is that superconductivity on the flat band Creutz lattice
is robust, $W_c\neq 0$, for all values of $U$. 

In the disorder-localized regime, it is important to determine the
disorder localization length, $\corrdisorder$, relative to the lattice
spacing and system size, since a short localization length would imply
strongly confined particles. On the other hand, even if the system is
insulating in the thermodynamic limit, it can have finite
superconducting transport on a finite lattice if the pairs maintain
phase coherence and the localization length is much larger than the
system size. We will, therefore, establish the functional dependence
and actual $\corrdisorder$ in the flat band systems by calculating the
pair correlation functions.

Our main results, obtained with large scale DMRG calculations, are as
follows. (1) At the two fermion densities we considered, (quarter and
eighth filling), the critical disorder required to destroy
superconductivity is finite, $W_c \neq 0$, for all values of
interaction strength we studied, $0 < U \leq 25$. For $U$ smaller than
the gap between the two flat bands ($U < 4t$) in the Creutz lattice,
the Luttinger parameter obeys $K > K_c = 3/2$ applicable to spinful
fermions with attractive interactions on a ladder. (2) For very large
$U$, the model is very well described by an effective HCB model on a
crossed ladder with $K_c = 3/4$ which is smaller than the $K$
calculated for our system. Consequently, the finite $W_c$ we find is
in agreement with bosonization predictions for dispersive bands. (3)
In the intermediate $U$ regime, where bosonization predictions are not
reliable, we find that the system superconductivity remains robust and
$W_c\neq 0$.  Consequently, superconductivity in this system is robust
and resists the localizing effect of disorder for all values of the
interaction $U$. (4) The critical disorder is directly proportional to
the SC weight, $D_s$, of the clean system at the same $U$:
$W_c(U)\propto D_s(U,W=0)$.

The paper is organized as follows. In Section \ref{sec:Model and
  Methods} we describe the Hubbard Hamiltonian with site disorder and
the flat band Creutz lattice that we consider. In Section \ref{sec:Results},
we first discuss the disorder localization length and its scaling
form. Thereafter, we study the effects of interaction strength on the
robustness of superconductivity in the presence of disorder. Lastly, we apply mean field theory (MFT) to the disordered system, and show the consistency between MFT and exact calculations for obtaining $W_c$. Our
conclusions are presented in Section \ref{sec:concs}.

\begin{figure}
    \centering
    \includegraphics[width=7.6cm]{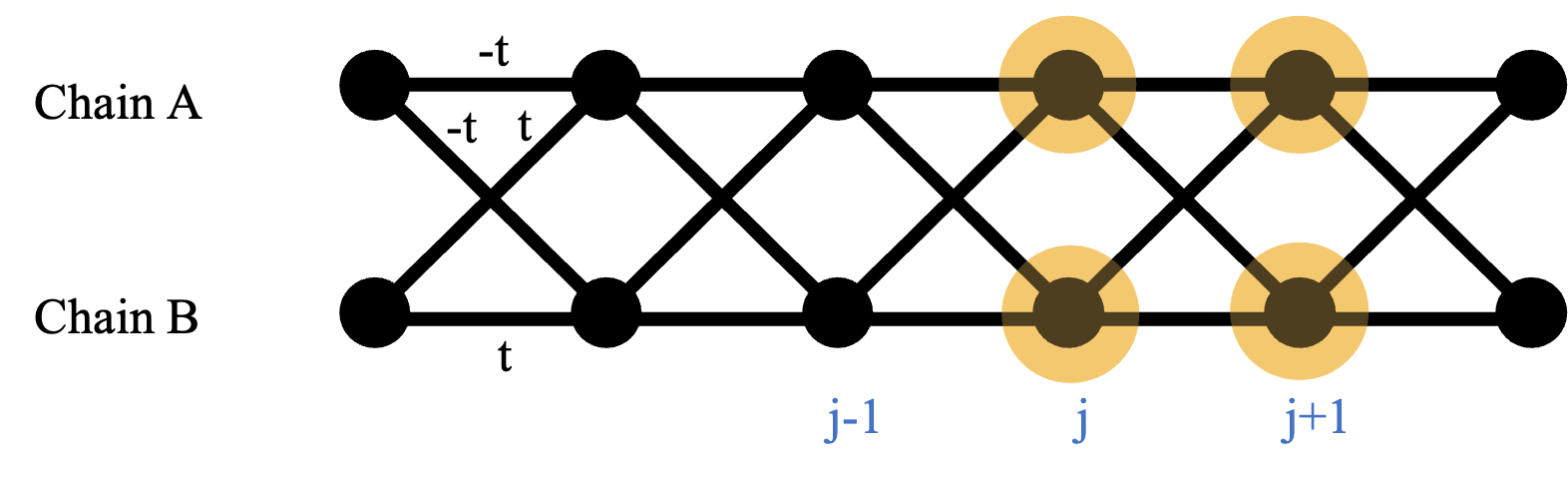} \\
    \includegraphics[width=7.6cm]{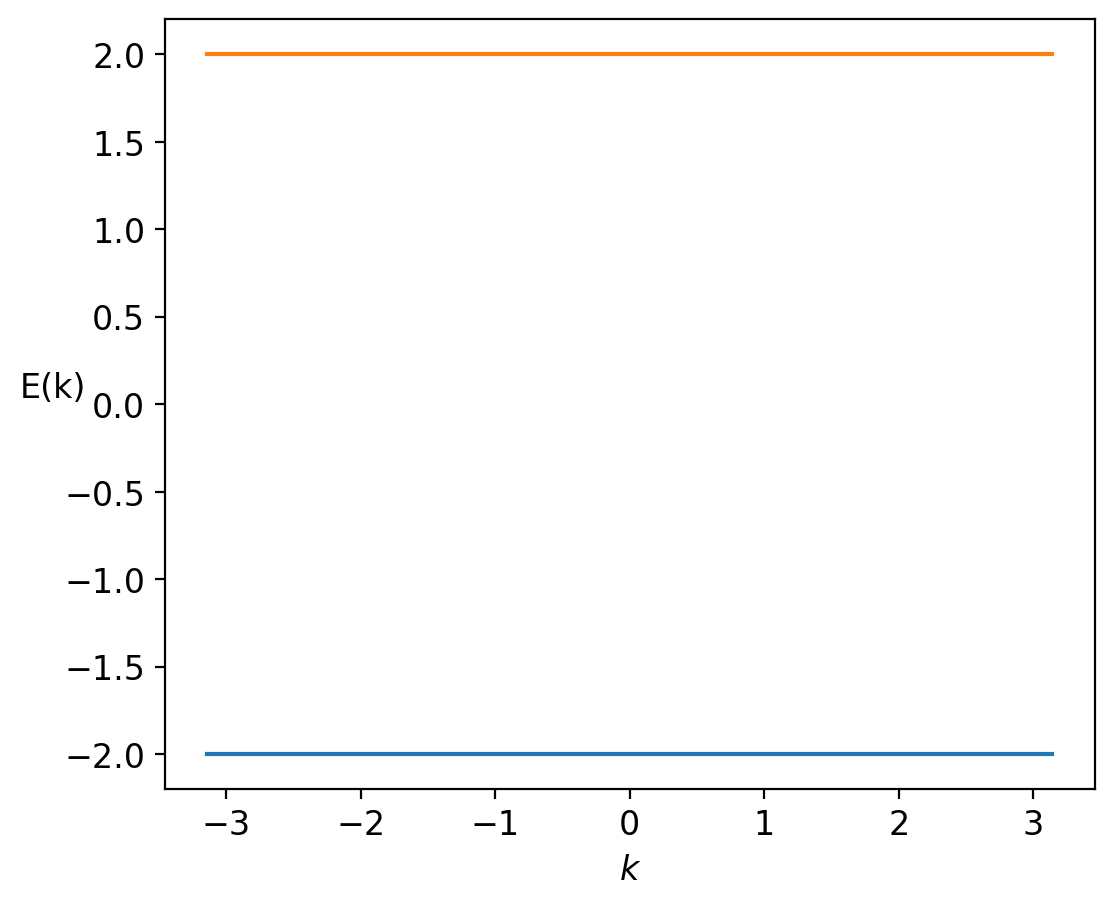}
    \caption{(Color online) \textbf{Top}: The Creutz lattice has two
      sublattices, with a (CLS) localized within two unit cells shown by
      the orange discs. \textbf{Bottom}: Band structure of the Creutz
      lattice with the chosen hopping potentials has two flat bands at
      $E(k)=-2$ and $2$.}
    \label{fig:CreutzLattice}
\end{figure}

\section{Model and Methods}\label{sec:Model and Methods}
In this study, we consider the Hubbard Hamiltonian with an additional
on-site disorder term. The Hubbard Hamiltonian with an attractive
on-site interaction is given by
\begin{equation}
\begin{aligned}
    H_\mathrm{Hubbard} = &\displaystyle\sum_{i,j,\alpha,\alpha^\prime,\sigma}\left(
    t_{ij}^{\alpha,\alpha'}c^{\alpha\dagger}_{i,\sigma}c^{\alpha'}_{j,\sigma}
    + h.c.\right) \\&-U \displaystyle\sum_{j,\alpha}
    c^{\alpha\dagger}_{j,\downarrow}
    c^{\alpha\dagger}_{j,\uparrow}c^{\alpha}_{j,\uparrow}c^{\alpha}_{j,\downarrow},
    \label{eq:hamilt}
\end{aligned}
\end{equation}
where $i,j$ are unit cell labels (running from $1$ to $L$), $\alpha,
\alpha'$ are sublattice indices, $t_{ij}^{\alpha,\alpha'}$ is the
hopping parameter between lattice sites ($i,\alpha$) and ($j,
\alpha'$), and $U>0$ is the strength of the attraction. The operator
$c^{\alpha}_{j,\sigma}$ ($c^{\alpha\dagger}_{j,\sigma}$) destroys
(creates) a spin $\sigma=\uparrow,\downarrow$ fermion on site $i,
\alpha$. With the diagonal disorder term, we have

\begin{equation}
    H = H_\mathrm{Hubbard}+\displaystyle\sum_{j,\sigma,\alpha}
    \mu_{j,\alpha} c^{\alpha\dagger}_{j,\sigma}c^{\alpha}_{j,\sigma}
\label{eq:hamiltdis}
\end{equation}
where $H_\mathrm{Hubbard}$ contains the hopping and Hubbard attraction
terms, and $\mu_{j,\alpha}\in[-W/2,W/2]$ is randomly drawn from an
uncorrelated, uniform distribution with width of the disorder strength
$W$.

We consider the Creutz lattice~\cite{creutz1999end} which is quasi
one-dimensional with two chains and can be described with the
inter-cell hopping Hamiltonian,
\begin{equation}
  \label{eq:Creutzham}
    H_1 = 
    \begin{pmatrix} -t &-t \\
    t& t
    \end{pmatrix}.
\end{equation}

With these hopping terms, the Creutz lattice has two flat bands,
$E(k)=-2t$, $2t$ in the non-interacting limit shown in
Fig.~\ref{fig:CreutzLattice}. The localized state is compact, with
$\ket{\Psi_j}=\tfrac{1}{2}(c_j^{A\dagger}+c_{j+1}^{A\dagger}-c_j^{B\dagger}
+c_{j+1}^{B\dagger})\ket{0}$
and equal filling on all sites of the compact localized state
(CLS). Additionally, it has a chiral symmetry and a winding number
$\mathcal{W}=1$ of the lower flat band.

We analyze the interplay between disorder and attractive interaction
and their effects on flat band superconductivity with DMRG and matrix
product state (MPS) optimization algorithms available in the ALPS
library~\cite{bauer2011alps} and the ITensor Software Library for Tensor Network Calculations~\cite{itensor, itensor-r0.3}. We also performed mean field theory
(MFT) calculations (see Sec.~\ref{sec:MFT}). To compute the superfluid
weight, $D_s$, we use DMRG with periodic boundary conditions (PBC) and
an applied phase gradient, $\phi\equiv \Phi/L$ where $\Phi$ is the
total phase twist applied on the system. This gradient is applied via
the transformation $c^\alpha_{j\sigma} \rightarrow {\rm e}^{{\rm
    i}\phi
  j}c^\alpha_{j\sigma}$~\cite{mondaini2018pairing,chan2022pairing}.
The superfluid weight is defined
by~\cite{fisher73,zotos90,shastry90,scalapino93,hayward95,batrouni2004metastable}
\begin{equation}\label{eq:Ds}
    D_s = \pi L \left.\frac{d^2 E_\mathrm{GS}(\Phi)}{d \Phi^2}\right\vert_{\Phi=0}
\end{equation}
in quasi $1D$. As shown in Ref.\cite{mondaini2018pairing}, in the pure
case (no disorder), $E_\mathrm{GS}(\Phi)$ is periodic in $\Phi$ with a
period of $\pi$, and $E_\mathrm{GS}(0<\Phi\leq \pi)$ is
quadratic. Therefore, in this case one can use the approximation $D_s
\approx \frac{8L}{\pi}(E_\text{GS}(\pi/2)-E_\text{GS}(0))$ where $E_\text{GS}$ needs
be determined at only two points. However, in the disordered case, we
find that the ground state energy, $E_\mathrm{GS}(\Phi)$, deviates
from quadratic behavior as $\Phi$ increases (see below).  We therefore
do not use this approximation; instead, we compute
$E_\mathrm{GS}(\Phi)$ for several small values of $\Phi$ ensuring that
they are within the quadratic region to find $D_s$. This is repeated
for multiple disorder realizations, up to chain lengths of $L=20$. The
typical number of states in the DMRG calculation is $1000$, and the
number of sweeps $20$. We verified that these choices yield results
with the desired precision.

To calculate the correlation functions and disorder localization
length, $\corrdisorder$, we use MPS optimization with open boundary
conditions (OBC) up to a system size of $L=200$ unit cells. Since the
mobile particles are bound pairs, the disorder localization length is
obtained through the pair Green's function (GF). In the absence of
disorder, the pair GF decays as a power law,
\begin{equation}
    \text{G}^{\alpha\alpha'}_{\text{pair}}(r)=\langle
    c_{j+r,\downarrow}^\alpha c_{j+r,\uparrow}^\alpha
    c_{j,\uparrow}^{\alpha'\dagger}c_{j,\downarrow}^{\alpha'\dagger}
    \rangle \sim r^{-\omega}=r^{-\frac{1}{2K}}
    \label{eq:pairGF}
\end{equation}
where $K$ is the Luttinger parameter. When disorder is introduced into
the system, and there is disorder-induced localization, the pair GF
decays exponentially
\begin{equation}
    \text{G}^{\alpha\alpha'}_{\text{pair}}(r)=\langle c_{j+r,\downarrow}^\alpha c_{j+r,\uparrow}^\alpha
    c_{j,\uparrow}^{\alpha'\dagger}c_{j,\downarrow}^{\alpha'\dagger}
    \rangle \sim {\rm e}^{-\frac{r}{\corrdisorder}}
\end{equation}
where $\corrdisorder$ is the disorder correlation length.

Using DMRG with PBC to compute $D_s$, we take up to 300 disorder
realizations, and for MPS with OBC to compute the Green's functions
and $\corrdisorder$, we take up to 40 realizations for each disorder
strength. With OBC, we obtain the correlation functions from the
middle of the system and calculate the pair GF towards both the left
and right ends of the chain.

We point out that at very large $U$ ($t/U \ll 1 $) the up and down
fermions are bound so tightly, the pair behaves as a hard core boson
and the system is very well described by the effective HCB
Hamiltonian~\cite{micnas1990},
\begin{eqnarray}
{\cal H}_{\rm eff} &=& -\frac{2t^2}{|U|}\sum_{j,\alpha,\beta} \big (
b^{\alpha\dagger}_jb^{\beta}_{j+1} + {\rm h.c.}\big )
\nonumber\\
&&+\frac{2t^2}{|U|}\sum_{j,\alpha,\beta} \big ( n^\alpha_j
n^\beta_{j+1} + n^\beta_{j+1}n^\alpha_j \big )\nonumber \\
&&-\sum_{i,\alpha} \mu_{j,\alpha} n_i^\alpha,
\label{eq:hardcoreham}
\end{eqnarray}
where $n^\alpha_j = b^{\alpha\dagger}_jb^\alpha_j$, and
$b^{\alpha\dagger}_j$ ($b^\alpha_j$) is a hardcore boson creation
(annihilation) operator on site $(j,\alpha)$. They satisfy
$\{b^\alpha_j,b^{\alpha\dagger}_j\}=1$, and
$[b^\alpha_j,b^{\beta\dagger}_r]=0$ for $j\neq r$ or
$\alpha\neq\beta$. Note that this effective HCB Hamiltonian is also defined on the Creutz lattice, but it has dispersive bands only, because all the hopping
terms have the same sign (see Fig.~\ref{fig:CreutzLattice} for the
hopping term signs which give the flat band).

To confirm the large $U$ limit of Eq.(\ref{eq:hamilt}), we also study $D_s$ and the correlation functions of the effective HCB model using ITensor \cite{itensor, itensor-r0.3}.

\begin{figure}
    \centering
    \includegraphics[width=8.3cm]{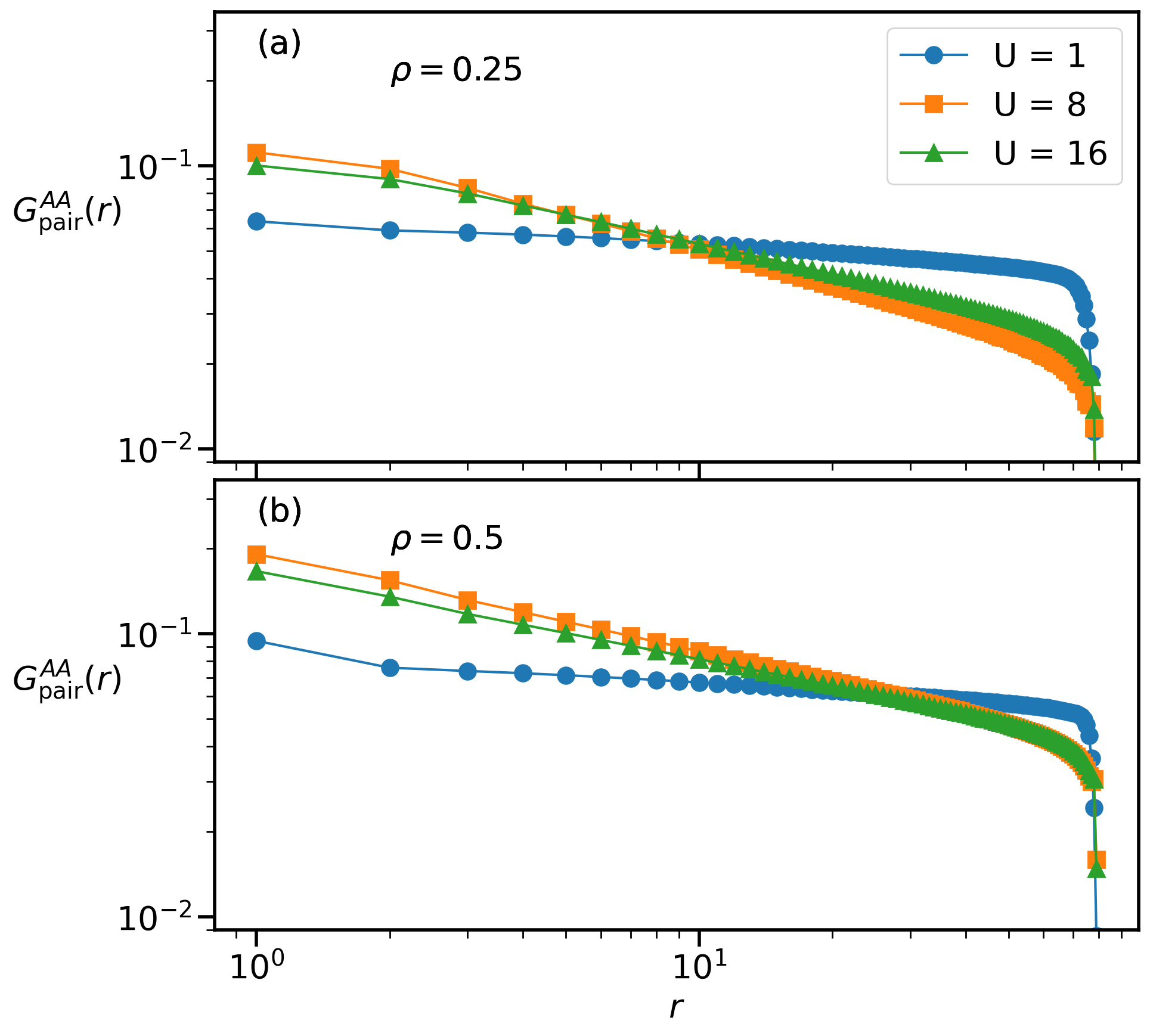}\\
        \includegraphics[width=8.6cm]{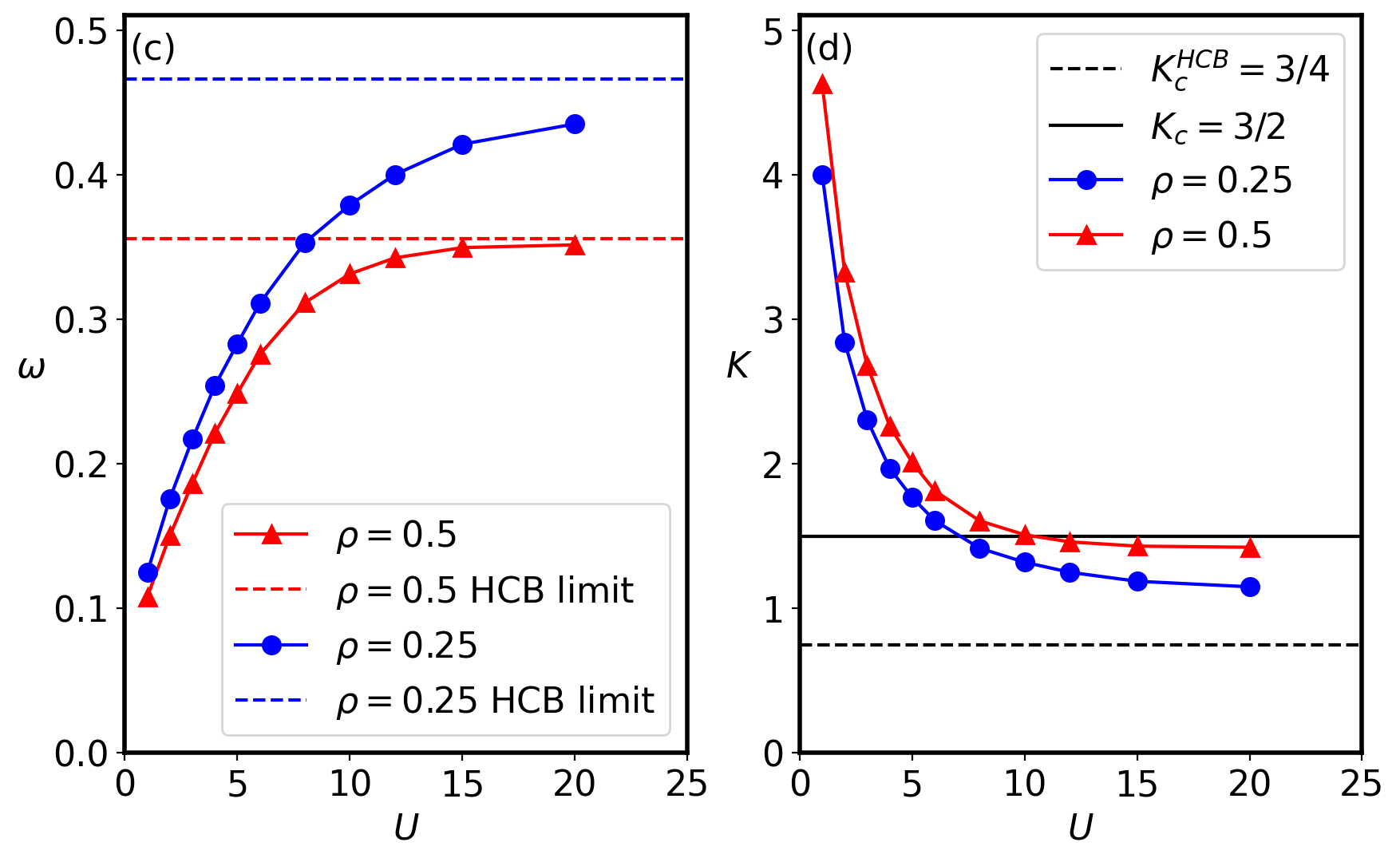}
    \caption{(Color online) (a) The pair correlation,
      Eq.(\ref{eq:pairGF}), for eighth filling, $\rho=0.25$, and (b)
      quarter filling, $\rho=0.5$. (c) The power law exponent,
      $\omega$, of the pair correlation function for
      $\rho=0.5,\,\,0.25$. The corresponding HCB exponents obtained
      with the effective model, Eq.(\ref{eq:hardcoreham}), are shown
      as horizontal dashed lines. (d) The correponding Luttinger
      parameters, $K=1/2\omega$. We also show the HCB critical value,
      $K^{\rm HCB}=3/4$ and the attractive fermion critical value,
      $K_c=3/2$.}
    \label{fig:Luttinger}
\end{figure}

\section{Results and Discussion}\label{sec:Results}
\subsection{Luttinger Parameters Revisited}

\begin{figure}
    \centering
    \includegraphics[width=8.6cm]{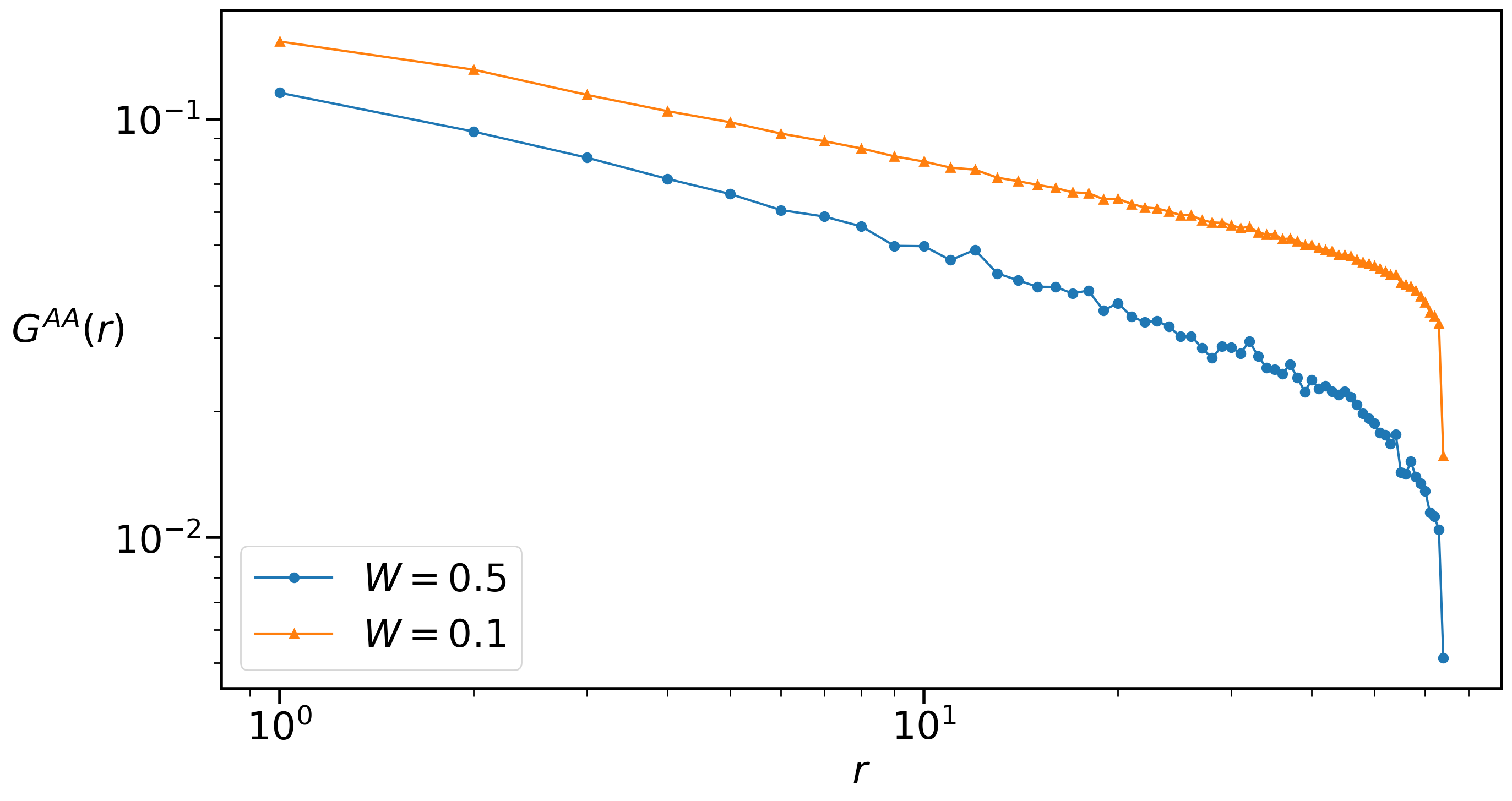}
    \caption{(Color online) Pair correlation functions at $\rho=0.5$, 
     $U=8$ for $L=128$ for disorder, $W=0.1$ and $W=0.5$. For
      $W=0.1$, the pair correlation function exhibits a clear power law behavior (i.e. a straight line in log-log plot), whereas for $W=0.5$, at intermediate distances, exhibits a curvature, emphasizing that it deviates from a power-law decay,  eventually becoming exponential at large distance. Note that the sharp decay at large distance is just a boundary effect and is pushed further away for larger lattice sizes). This emphasizes that a finite amount of disorder is
      needed to localize the system,  in agreement with a critical disorder strength $W_c=0.272$ extracted from the collapse of the superfluid weight. }
          \label{fig:Fermcorrfct}
\end{figure}

We revisit the Luttinger parameters in the case of the pure system to
confirm the values found in Ref.\cite{mondaini2018pairing} and to set
the stage to interpret our localization results in view of
bosonization predictions. Figure~\ref{fig:Luttinger}(a-b) show, in the
absence of disorder, the power law behavior of the pair GF,
Eq.(\ref{eq:pairGF}), for two fillings, $\rho=0.5$ (quarter filling)
and $\rho=0.25$ (eighth filling), at several values of $U$. Here, we define $\rho = \langle
c^\dagger_{j\uparrow}c^{\phantom\dagger}_{j\uparrow} + c^\dagger_{j\downarrow}c^{\phantom\dagger}_{j\downarrow}\rangle$ and $t=1$. The pair
GF yields the power law exponent, $\omega$, which is plotted in
Fig.~\ref{fig:Luttinger}(c) versus the coupling $U$. In
Fig.~\ref{fig:Luttinger}(d) we show the corresponding Luttinger
parameter $K=1/2\omega$.

\begin{figure}
    \centering
    \includegraphics[width=8.6cm]{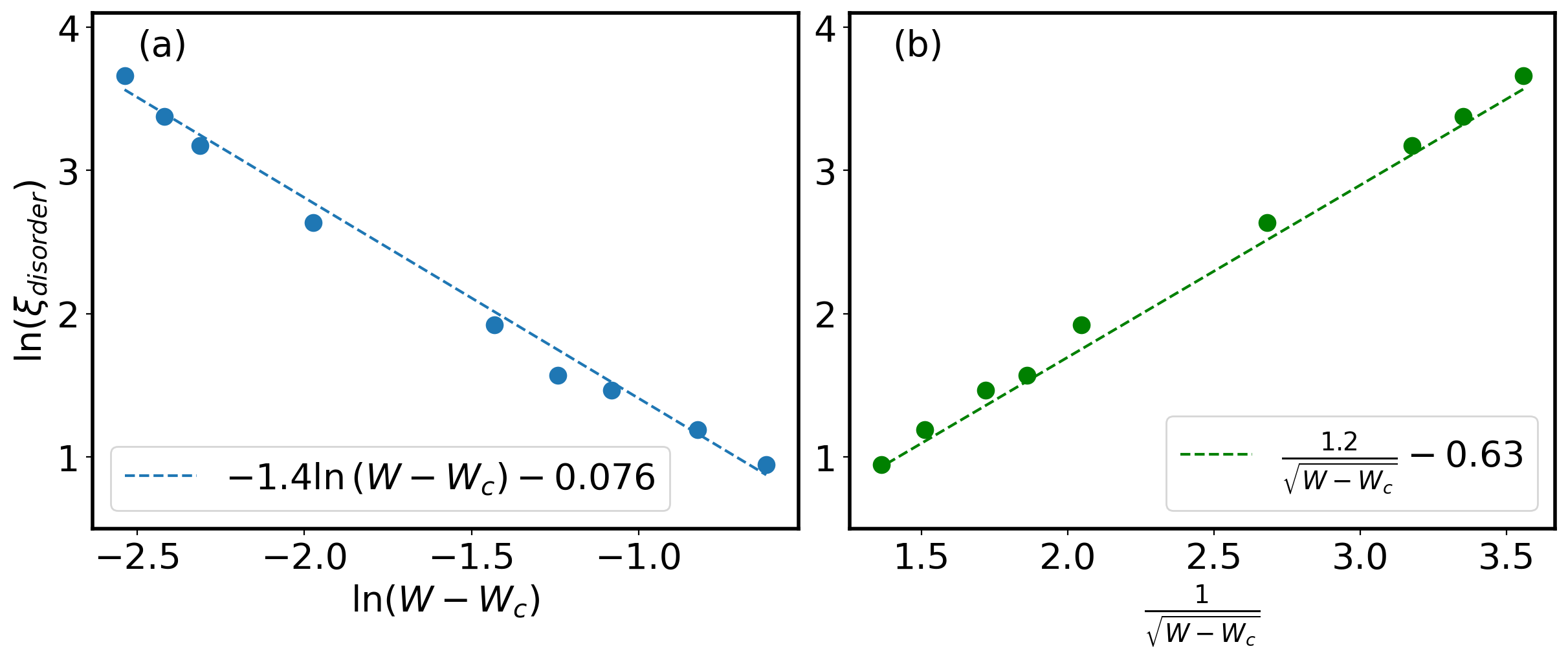}\\
        \includegraphics[width=8.6cm]{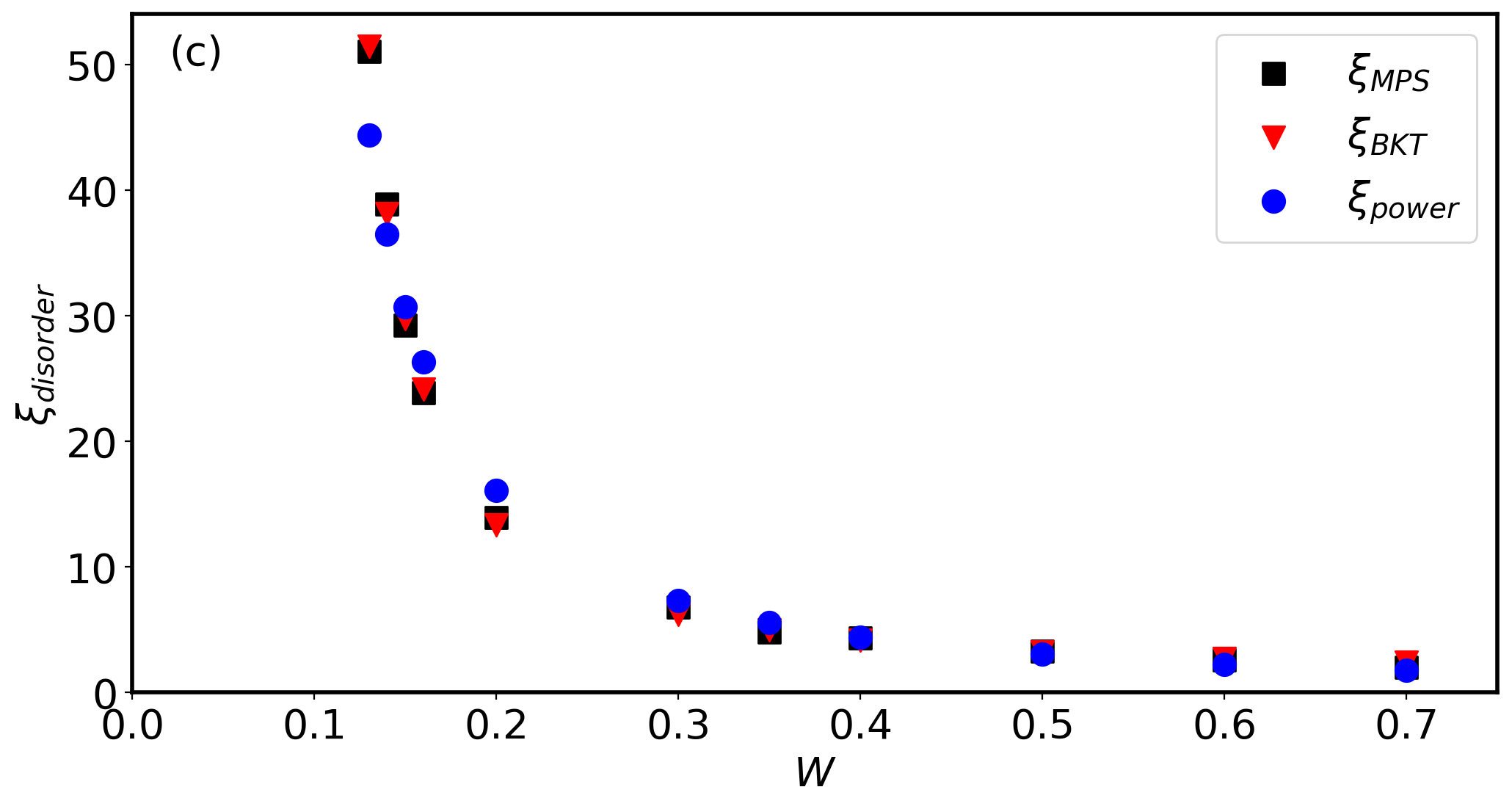}
    \caption{(Color online) Creutz lattice, $\rho = 0.5$,
      $U=1$. $\ln(\corrdisorder)$ plotted as a function of
      $\ln(W-W_c)$ (a) and $\frac{1}{\sqrt{W-W_c}}$ (b). The plot for
      $\frac{1}{\sqrt{W-W_c}}$ shows a better fit, agreeing with the BKT-type
      scaling form (Eq.~\eqref{eq:BKTform}). Here, $W_c=0.061$. (c) The
      disorder correlation length $\corrdisorder$ as a function of $W$
      computed using the power law ($\xi_{power}$) and BKT form
      ($\xi_{BKT}$), as well as direct measurement obtained using MPS
      optimization ($\xi_{MPS}$). $\xi_{MPS}$ agrees best with
      $\xi_{BKT}$. The localization length was computed on
        systems of up to $L=200$ and open boundary conditions.
      }
          \label{fig:CreutzDisorderCorrScaling}
\end{figure}

We mentioned above that in the limit $t/U\ll 1$, the effective HCB
model, Eq.(\ref{eq:hardcoreham}), describes the system very well. In
Refs.\cite{mondaini2018pairing,chan2022pairing,chan2022designer} this
was demonstrated by calculating the superfluid weight, $D_s$, and
showing that in the large $U$ limit it agrees very well with $D_s$
calculated with the original Hamiltonian, Eq.(\ref{eq:hamilt}). Here,
we add to this confirmation by calculating the power law exponent of
the HCB correlation function $\langle b_j^{\alpha} b_i^{\alpha \,
  \dagger}\rangle \propto |j-i|^{-\omega^\prime}$. Recall that a hard
core boson represents a fermion {\it pair} in the very strong
interaction limit, so a density of $\rho = 0.25\,\, (0.125)$ for HCB
is to be compared with a density of $\rho=0.5\,\, (0.25)$ for
fermions. For HCB $\rho = 0.25$ (compare with fermion density $0.5$)
we find $\omega^\prime = 1/(2K) = 0.354$ while for $\rho=0.125$
(compare with fermion density $0.25$ we find $\omega^\prime = 1/(2K) =
0.467$. These HCB exponents are shown in Fig.~\ref{fig:Luttinger}(c)
as horizontal dashed lines demonstrating the asymptotic approach to
these values of the fermion pair GF power law exponents.

Figure~\ref{fig:Luttinger}(d) shows for both $\rho=0.5$ and
$\rho=0.25$, that the Luttinger parameter at lower $U$ values, where
bosonization is reliable, is larger than $K_c$ with the consequence
that the critical localizing disorder strength for this range of $U$
values, should be finite, $W_c\neq 0$. In addition, we see that for
large $U$, as the system approaches the HCB limit, $K$ is always
larger than $K_c^{\rm HCB}$, again implying that $W_c$ should have a finite nonzero value. The finiteness of $W_c$ and sustained superconducting state will be demonstrated below by direct computations.

\begin{figure}
    \centering
    \includegraphics[width=8cm]{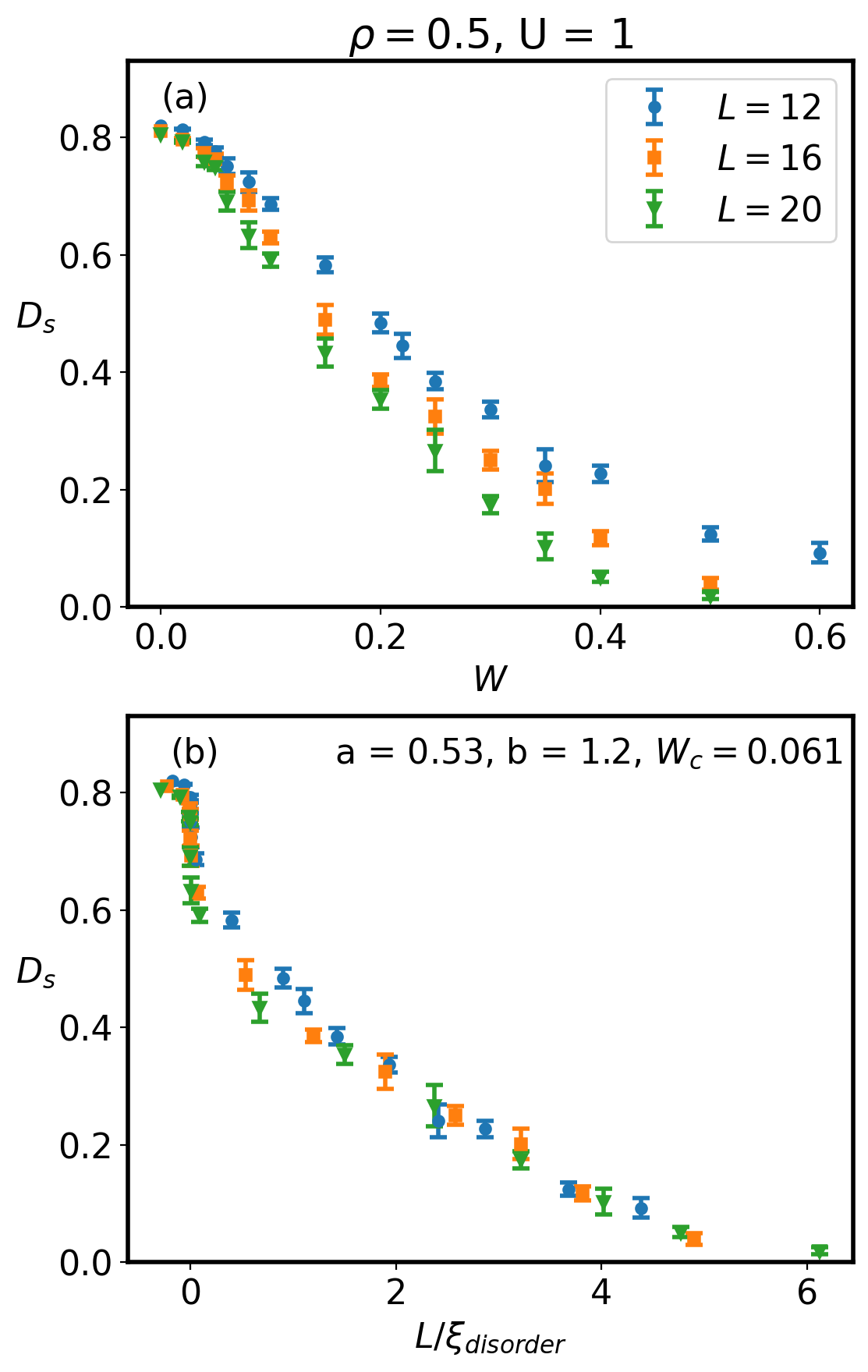}
    \caption{(Color online) (a) Superfluid weight as a function of
      disorder strength on the Creutz lattice with $U=1$ and
      $\rho=0.5$. (b) Using the values of $\corrdisorder$ obtained
      from the BKT scaling in
      Fig.~\ref{fig:CreutzDisorderCorrScaling}, we find the best
      collapse of $D_s$ as a function of $L/\corrdisorder$ for $W_c=
      0.061$.}
    \label{fig:CreutzDisorderU1}
\end{figure}

\subsection{Disorder Scaling Form}\label{sec:CreutzU1}

We first discuss the scaling form of the disorder localization length,
$\corrdisorder$, that should be used in the many-body localization
transition. The disorder localization length, or disorder correlation
length, $\corrdisorder$, measures how far the particle can travel
before a significant decay in the wavefunction. In the presence of
disorder but without the fermion-fermion on-site interaction, it is
well known that in quasi $1D$ the scaling of the localization length
follows a power law with an exponent of $2$ for dispersive
bands~\cite{kramer1993localization,chayes1986finite} and, for flat
bands, an exponent that depends on the class of flat
band~\cite{leykam2017localization,leykam2013flat}. However, in the
many-body system, \textit{i.e.} in the presence of the fermion-fermion
attractive interaction, there have been studies of the scaling form of
$\corrdisorder$ near the critical point for spin models, without a
flat band~\cite{vsuntajs2020ergodicity, vsuntajs2020quantum}. These
were determined by minimizing the cost function (see Appendix
\ref{appendix:costfunc}) and scaling the relevant parameters (in their
case, the eigenstate entanglement entropy, level spacing ratio and the
ergocity indicator). The candidate scaling functions are the BKT and
the power law forms:

\begin{eqnarray}
  \corrdisorder&=&a
  \label{eq:BKTform}
  e^{\frac{b}{\sqrt{\abs{W-W_c}}}}=a\corrdisorder^*\\
  \label{eq:powerform}
  \corrdisorder&=&a'\abs{W-W_c}^{-\mathit{n}}=a'\corrdisorder^*
\end{eqnarray}
where $a$, $b$, $W_c$, $a'$, and $\mathit{n}$ are constants to be
determined. Eq.~\eqref{eq:BKTform} is the BKT scaling form while
Eq.\eqref{eq:powerform} is the power law form of a second-order transition.

In order to determine which of these two scaling forms to use, we
calculate $\corrdisorder$ directly on long chains and examine how it
diverges as a function of $W$, the disorder strength.

We obtain $\corrdisorder$ by computing the pair GF with MPS
optimization on systems of lengths (\textit{i.e.} number of unit
cells) of up to $L=200$ and open boundary conditions. An example of
this correlation function is shown in Fig.~\ref{fig:Fermcorrfct} for
$\rho=0.5$, $U=8$ and  $L=128$. The pair correlation function exhibits a clear power law behavior (i.e. a straight line in log-log plot), whereas for $W=0.5$, at intermediate distances, exhibits a curvature, emphasizing that it deviates from a power-law decay,  eventually becoming exponential at large distance. Note that the sharp decay at large distance is just a boundary effect and is pushed further away for larger lattice sizes). This behaviour is in agreement with the critical value $W_c=0.272$ extracted from the collapse of the superfluid density. This emphasizes that there is a critical value of
disorder needed to localize the pairs. In
Fig.~\ref{fig:CreutzDisorderCorrScaling}(a) and (b), we plot the
logarithm of the disorder correlation length ($\ln(\corrdisorder)$) as
a function of $\ln(W-W_c)$ and $1/\sqrt{W-W_c}$, where in both cases we
optimize the cost function to determine the fitting parameters. We see
that the BKT form gives a better fit and yields $W_c=0.061$. This
confirms that the proper choice is the BKT form
(Eq.~\eqref{eq:BKTform}) as was used in
Ref.~\cite{Liang2023Disorder}. We also show in
Fig.~\ref{fig:CreutzDisorderCorrScaling}(c) $\corrdisorder$ computed
with the two fits and the data measured from MPS optimization. At
strong disorder, both scaling forms show reasonable agreement with
exact MPS optimization data, but as we tune to smaller values of
disorder, it is evident that the power law scaling does not describe
the actual behavior of $\corrdisorder$. Moreover, one must consider
the parameters $a$ and $a'$, to calculate the value of $\corrdisorder$
accurately. This is crucial to obtain the actual length scale of the
disorder localization, and not simply as a functional scaling.

To confirm the validity of this fit, we calculate the superfluid
weight, $D_s$, on the Creutz lattice as a function of disorder
strength, $W$, with PBC and $U=1$ (Fig.~\ref{fig:CreutzDisorderU1}(a))
for system sizes of $L=12$, $16$ and $20$. In
Fig.~\ref{fig:CreutzDisorderU1}(b), we scale these data as a function
of $L/\corrdisorder$, where $\corrdisorder=0.53{\rm
  e}^{1.2/\sqrt{W-0.061}}$ is obtained from the direct measurement of
$\corrdisorder$ in Fig.~\ref{fig:CreutzDisorderCorrScaling}. The
collapse of the superfluid weight (computed with DMRG on a PBC
lattice) as a function of $L/\corrdisorder$ (measured with MPS and
OBC) is very good, thus confirming the consistency of the two distinct
approaches.

\begin{figure}
    \centering
    \includegraphics[width=8cm]{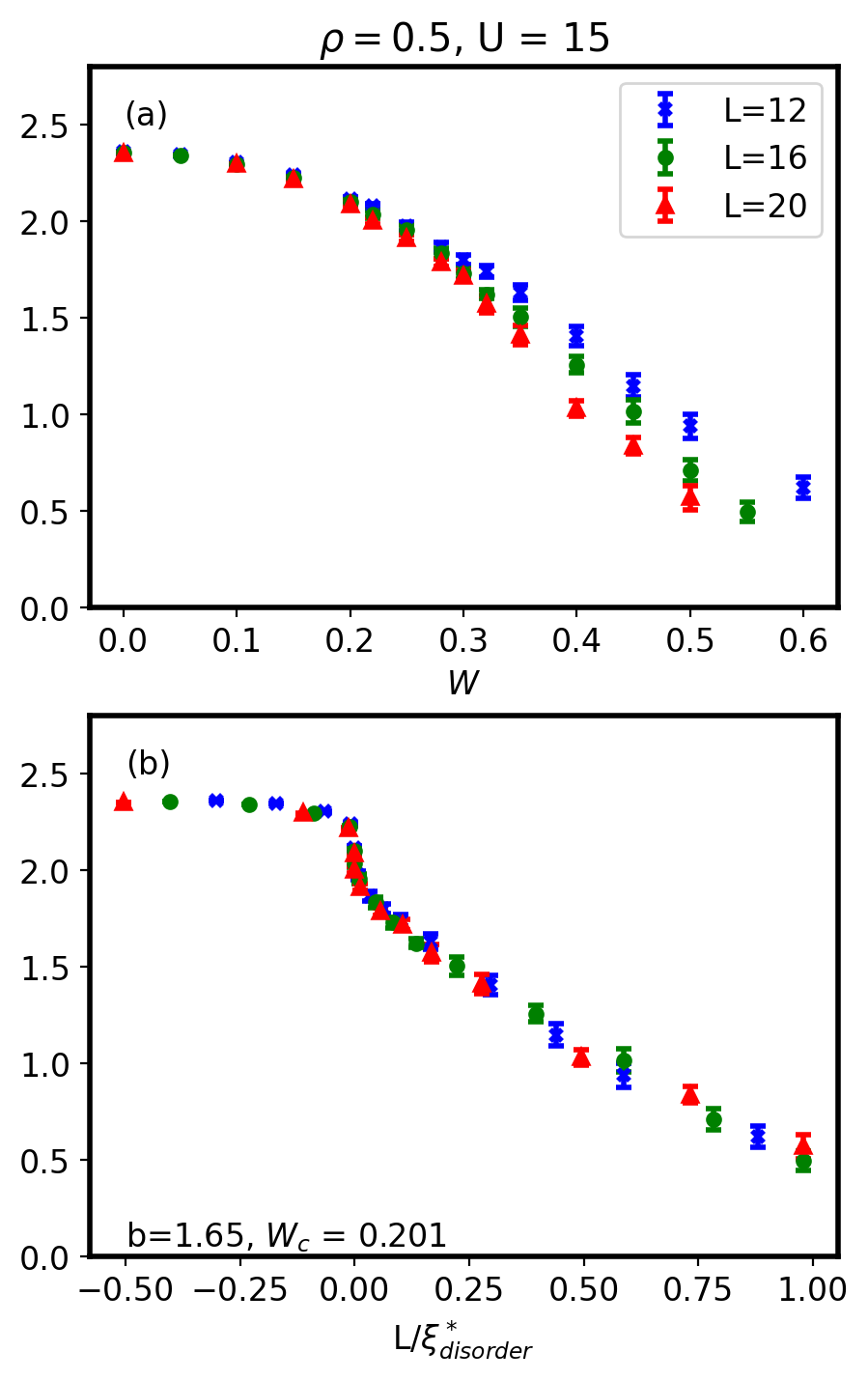}
    \caption{(Color online) (a) Superfluid weight as a function of the
      disorder strength on the Creutz lattice with $U = 15$. By
      minimizing the cost function, we plot $D_s$ as a function of
      $L/\xi^*_\text{disorder}$; the critical disorder is
      $W_c=0.2$. $\rho = 0.5$.}
    \label{fig:CreutzDisorderU15}
\end{figure}

Using MPS optimization, we showed that the scaling form and prefactor $a$
cannot simply be extracted from minimizing the cost function for $D_s$
at various system sizes. It is important to know accurately the length
scale of $\corrdisorder$, because if the system is small compared with
$\corrdisorder$, the pair coherence length would exceed the system
size with the consequence that vestiges of SC remain in the system.
However, if one were only concerned with finding the critical
disorder, $W_c$, and has determined the scaling form, then minimizing
the cost function remains a judicious approach.

\subsection{Interplay between Interaction and Disorder}

We now study the effect of disorder on the superconducting phase. To
this end, we calculate the superfluid weight, $D_s$, Eq.(\ref{eq:Ds}),
as a function of disorder strength, $W$, for several system sizes,
$L$. Fig.~\ref{fig:CreutzDisorderU1}(a) shows this for $U=1$. We see
that as $W$ increases, $D_s$ decreases, and as $L$ is increased, $D_s$
decreases even faster with disorder strength for sufficiently large $W$. We perform finite size scaling by plotting $D_s$ versus the scaling
variable $L/\xi_{\rm disorder}$ with $\xi_{\rm disorder}$ given by the
BKT form, Eq.(\ref{eq:BKTform}). In this equation, the parameters $b$
and $W_c$ are tuned to minimize the fitting cost function (Appendix
\ref{appendix:costfunc}) and obtain the values of $b$ and $W_c$ which
give the best data collapse in the critical region. The collapsed data
are shown in Fig.~\ref{fig:CreutzDisorderU1}(b) and give a nonzero
critical disorder value, $W_c = 0.061$. Bosonization of spinful
fermions with attractive interaction and dispersive bands on a
ladder~\cite{orignac1996effects,orignac1997effects,orignac1999}
predict that if the Luttinger parameter, $K$, exceeds a critical
value, $K_c= 3/2$, the system will require the disorder strength to
exceed a critical value, $W>W_c$, in order to be localized. Here we
found that, at $U=1$, our system requires $W> W_c=0.061$ to be
localized. Fig.~\ref{fig:Luttinger}(c) shows for $\rho=0.5$ at
$U=1$, the power law decay exponent $\omega = 1/2K \approx 0.11$,
giving $K\approx 4.55$ which exceeds $K_c$. So, assuming that the
bosonization prediction also applies to this flat band ladder system,
our result of finite $W_c$ for $U=1$ and bosonization are in
agreement. In fact, Fig.~\ref{fig:Luttinger}(d) shows that for
$\rho=0.5$, $K>K_c$ for all $U\lesssim 10$. Bosoniztion, therefore,
predicts that $W_c\neq 0$ for all $U\lesssim 10$.

\begin{figure}
    \centering
    \includegraphics[width=8cm]{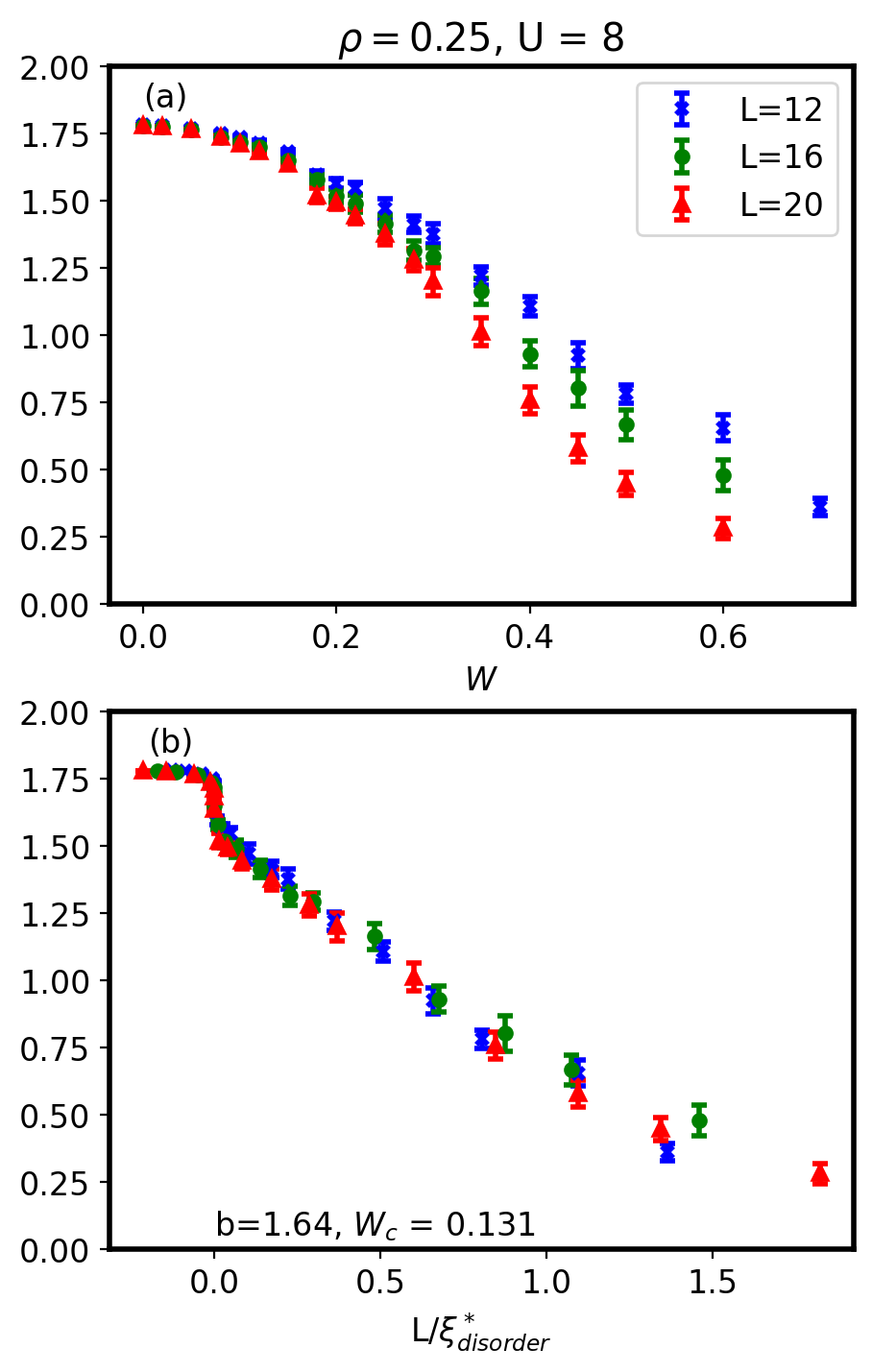}
    \caption{(Color online) (a) Superfluid weight as a function of the
      disorder strength on the Creutz lattice with $U = 8$. By
      minimizing the cost function, we plot $D_s$ as a function of
      $L/\xi^*_\text{disorder}$; the critical disorder is
      $W_c=0.131$. Here, $\rho = 0.25$.}
    \label{fig:CreutzDisorderU8rho0.25}
\end{figure}

\begin{figure}
  \centering
  \includegraphics[width=8.6cm]{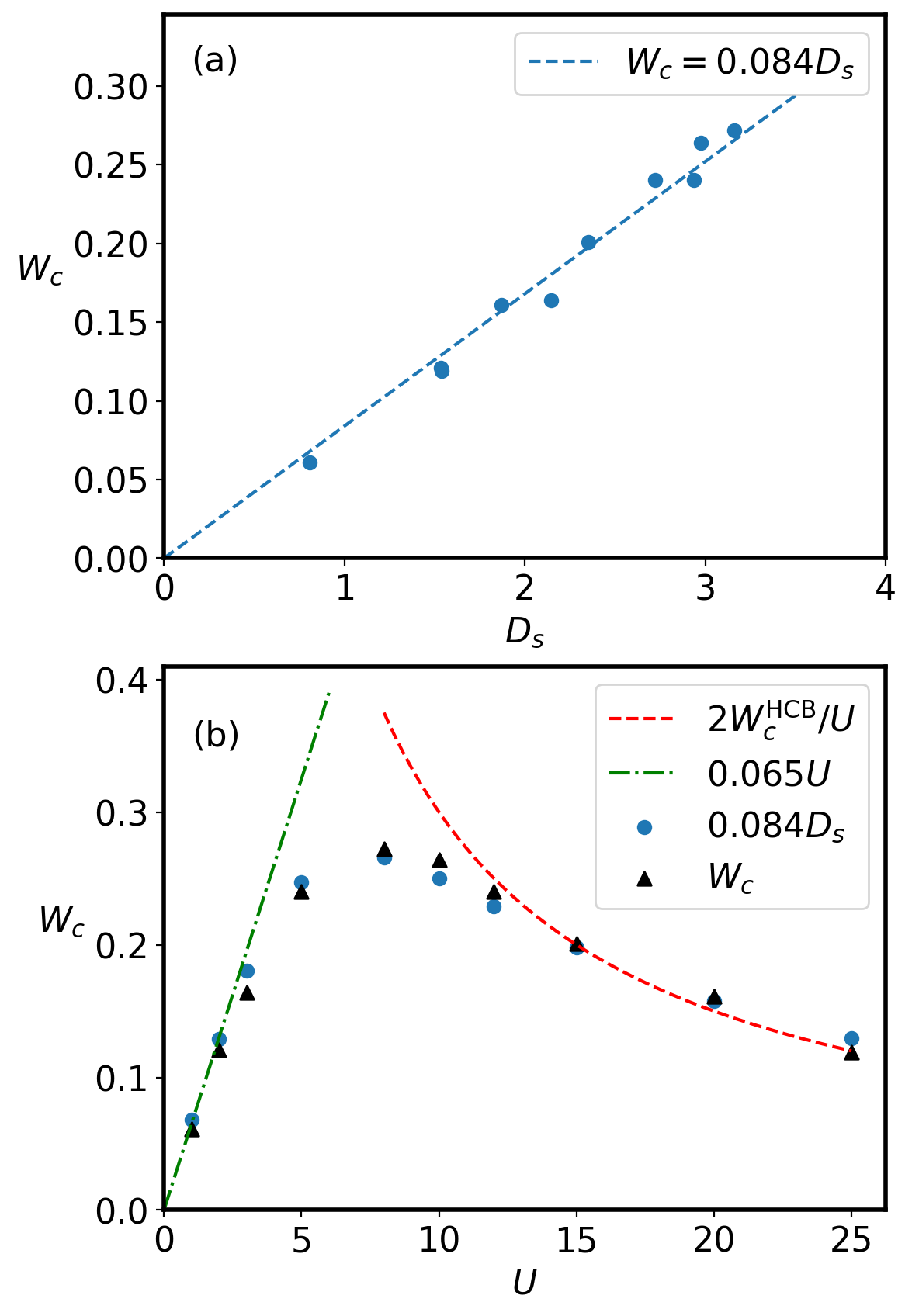}
    \caption{(Color online) (a) $W_c$ vs $D_s$ showing a very good
      linear dependence. (b) $W_c$ vs U for spinful
      fermions (black triangles) at $\rho=0.5$ obtained with the data collapse method
      illustrated in Figs. \ref{fig:CreutzDisorderU1},
      \ref{fig:CreutzDisorderU15} and
      \ref{fig:CreutzDisorderU8rho0.25}. At weak attraction, the
      dependence of $W_c$ on $U$ is linear, shown by the green 
      dash-dot line. The red dashed line represents $2W^{\rm HCB}_c/U$
      for the effective HCB model (at $\rho_\text{HCB}=0.25$) (see
      Fig.~\ref{fig:HCBDisorderrho0.25}). The
      data agreement is excellent between $Wc$ and $0.084D_s$. }
    \label{fig:WcDs}
\end{figure}

\begin{figure}
    \centering
    \includegraphics[width=8.5cm]{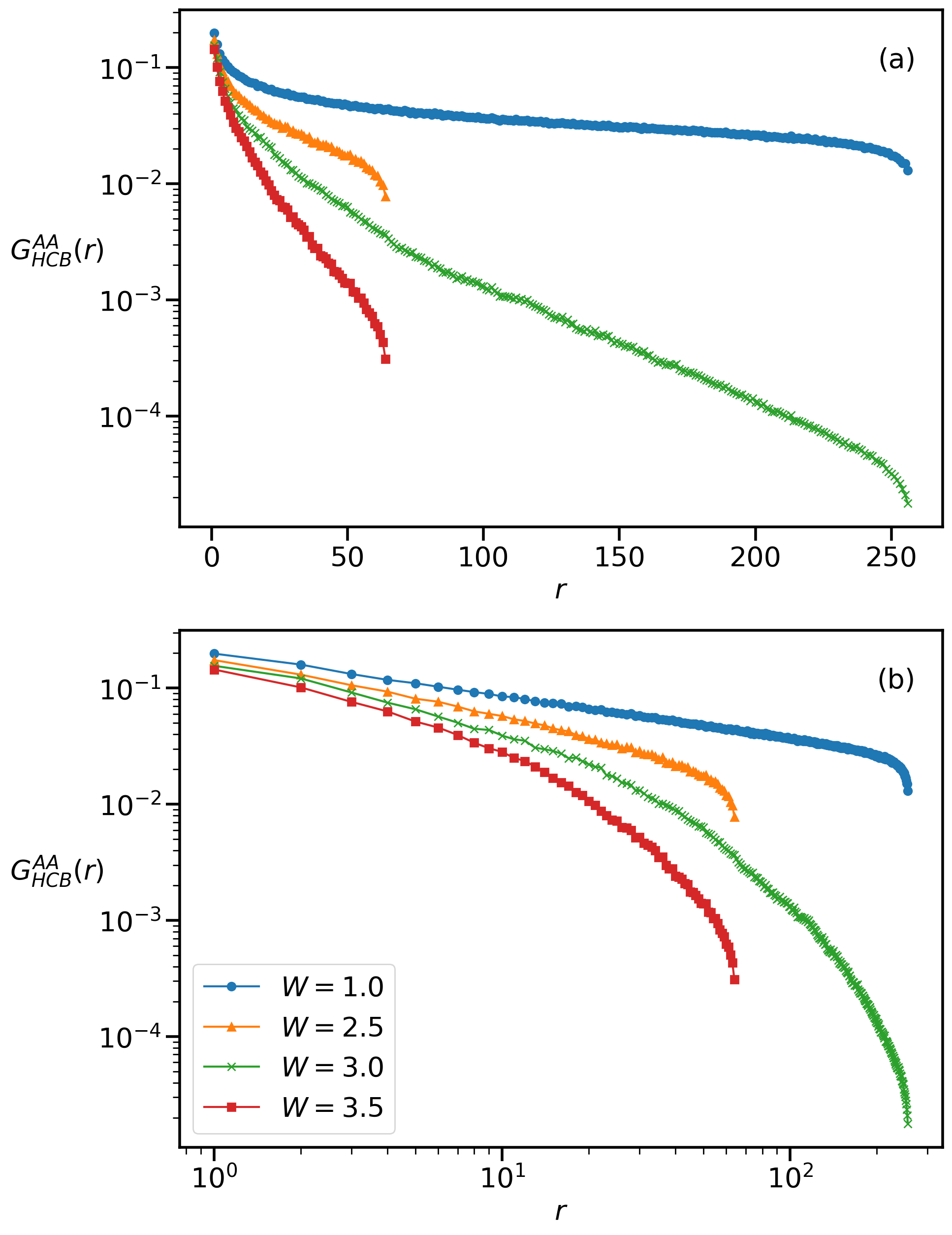}\\
    \caption{(Color online) The correlation functions in the effective
      HCB model, Eq.~\eqref{eq:hardcoreham} for $\rho_{\rm HCB}=0.25$
      and several values of disorder; (a) Semi-log scale and (b)
      log-log scale. The decay is a power law for $W=1$ and
      exponential for the other values showing that $W_c^{\rm HCB}$ is
      finite.  }
    \label{fig:HCBcorrfct}
\end{figure}

Fig.~\ref{fig:Luttinger}(c) also shows that as $U$ increases, so
does $\omega$ saturating at $\omega= 0.354$ (the $U\to \infty$ value)
with a Luttinger parameter value $K=1.41 < K_c$,
Fig.~\ref{fig:Luttinger}(d). Since for large $U$ we have $K<K_c$, the
question arises as to whether the fermion pairs are localized by any
amount of disorder, no matter how small. Figure
\ref{fig:CreutzDisorderU15} shows the same as
Fig.~\ref{fig:CreutzDisorderU1} but for $U=15$, where $\omega = 0.353$
and $K=1.4 < K_c=1.5$. In spite of the fact that $K<K_c$, we see that
at $U=15$, finite disorder strength is still required to localize the
system. This might suggest that this result disagress with
bosonization. However, we emphasize that at these very large values of
$U$, the paired fermions behave as very tightly bound hard core bosons
described by the effective Hamiltonian
Eq.~\eqref{eq:hardcoreham}. Therefore in this range of large $U$ values,
one must not use $K_c$ for spinful fermions with attractive
interactions, one must use instead the critical $K$ for hardcore
bosons on a ladder: $K^{\rm HCB}_c =
3/4$.\cite{crepin2011phase}.  Then, at $U=15$, $K=1.4 > K_c^{\rm HCB}$ and the system is
localized only if the disorder strength exceeds a certain finite
critical value, $W_c \neq 0$.

To confirm that at large $U$, where $K$ is smaller than the $K_c$ for
spinful fermions with attraction on a ladder, the system still
requires finite disorder to localize, we also examine the $\rho=0.25$
case where $\omega(U\to \infty) =0.467$ yielding $K=1.07$ well below
$K_c=3/2$. Fig.~\ref{fig:CreutzDisorderU8rho0.25} shows that at $U=8$, finite disorder strength, $W_c = 0.131$, is required for
localization. This is further evidence that at large $U$, $K_c$ is not
given by the spinful fermion case, but by the HCB value, $K_c^{\rm HCB}$.

Using the methods that produced Figs.
~\ref{fig:CreutzDisorderU15}, and \ref{fig:CreutzDisorderU8rho0.25} we
determine $W_c$ as a function of $U$ at $\rho=0.5$ for $0<U\leq 25$,
and show the results in Fig.~\ref{fig:WcDs}(b). This $W_c$ vs $U$
curve has the same general shape as $D_s$ vs $U$ in the clean
system\cite{mondaini2018pairing,chan2022pairing}. In particular, $W_c$ is linear in $U$
for small values, as is $D_s$ in the clean case.  We therefore plot in
Fig.~\ref{fig:WcDs}(a) $W_c$ as a function of $D_s$ at the same $U$
found for the pure system. We find that $W_c \propto D_s$ with the
proportionality constant $0.084$. Figure
\ref{fig:WcDs}(b) shows $W_c$ and $0.084D_s$ vs $U$ and illsutrates
this proportionality. We therefore argue that in such quasi
one-dimensional flat band systems, $W_c$ is finite and proportional to
$D_s$, the superfluid weight found for the pure system. Note that the
slope of $W_c$ as a function of $U$ is $0.065 = 0.084 * \pi
U\rho(1-\rho)$, where $D_s = \pi U\rho(1-\rho)$ at weak $U$ as shown
in Ref.~\cite{mondaini2018pairing}.

\begin{figure}
    \centering
    \includegraphics[width=8cm]{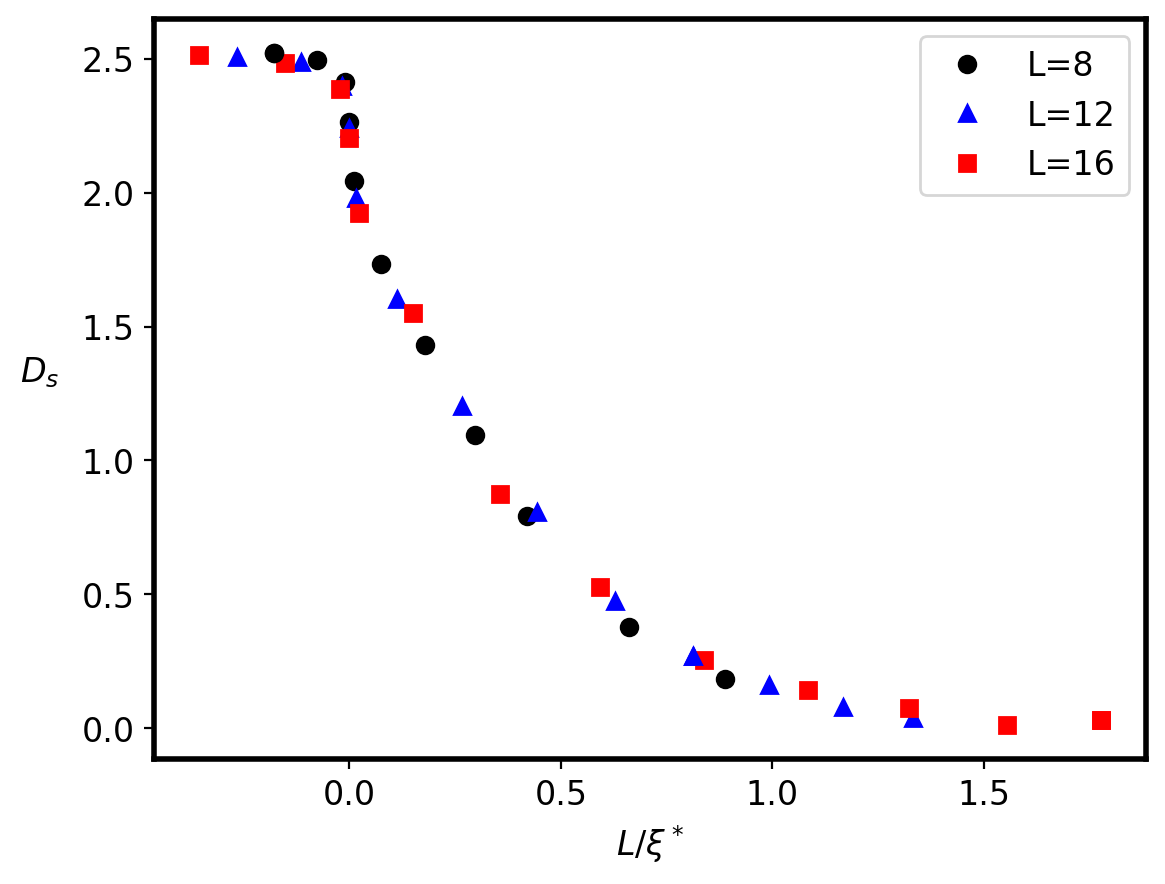}
    \caption{(Color online) Data collapse for the HCB effective
      model. Since a hard core boson represents a pair, the
      fermionic system at $\rho=0.5$ is represented by the HCB model
      with $\rho_\text{HCB}=0.25$. $W^{\rm HCB}_c = 1.5$.}
    \label{fig:HCBDisorderrho0.25}
\end{figure}

\begin{figure}
    \centering
    \includegraphics[width=7.8cm]{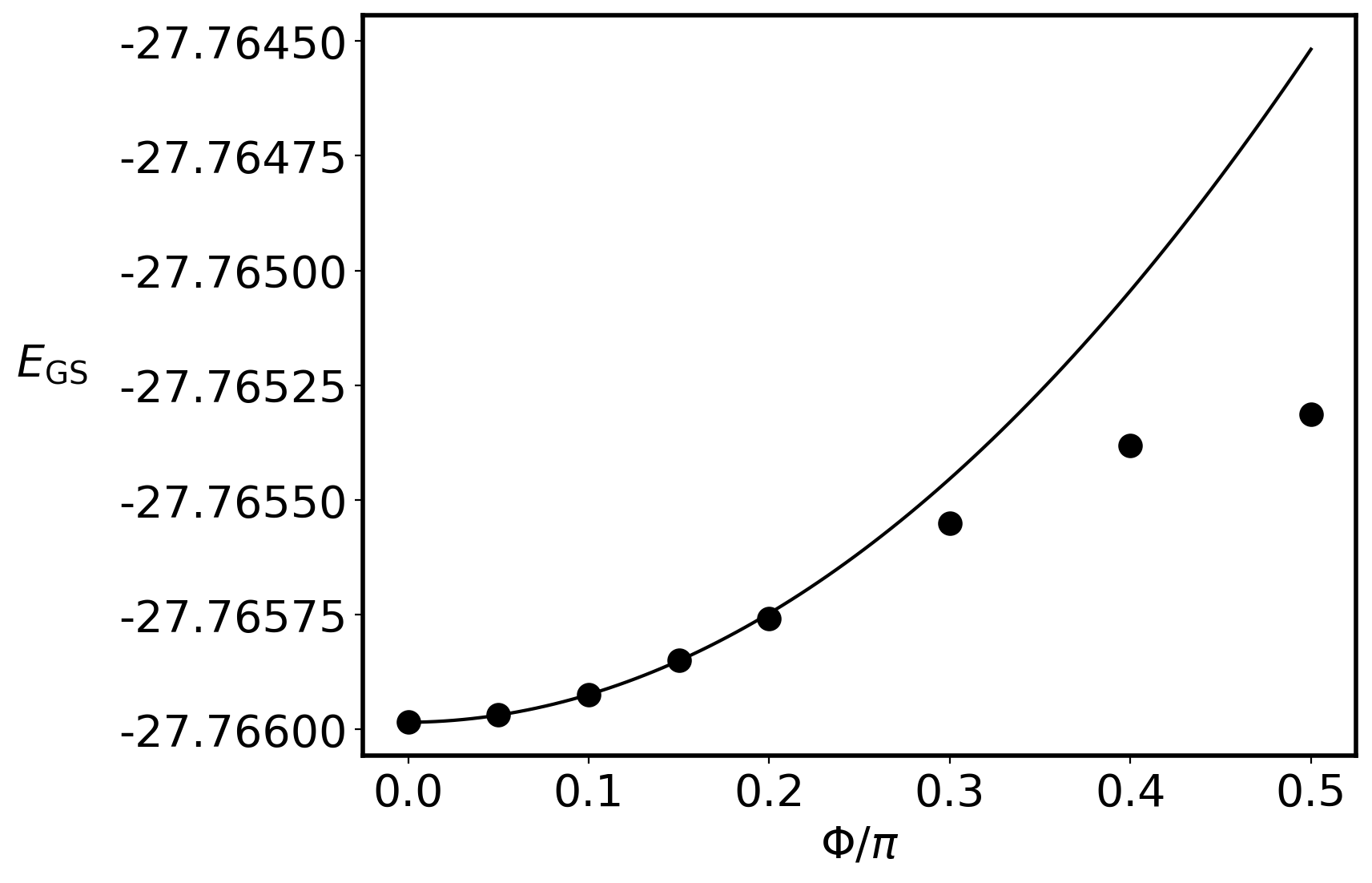}
    \caption{An example of the quadratic fit used to calculate $D_s$
      and points for $E_\mathrm{GS}(\Phi)$ up to $\Phi=0.5\pi$. In the
      presence of disorder, $E_\mathrm{GS}$ deviates from the
      quadratic behavior in $\Phi$, unlike the pure case where it
      remains quadratic for the entire range of
      $-\pi/2<\Phi<\pi/2$. (Creutz Lattice, $L=12$, $\rho=0.5$ , $U=1$ and
      $W=0.5$)}
    \label{fig:EG_Phi_Disorder}
\end{figure}
To study the role played by hard core boson physics in the large $U$
regime, we used DMRG with the HCB effective model,
Eq.~\eqref{eq:hardcoreham}. In this DMRG computation, $D_s$ was
obtained by keeping $1800$ states and doing $128$ sweeps. For the
correlations functions on $L=128,\,256$ we kept $2400$ states and
$128$ sweeps while for $L=512$, we did $128$ sweeps and kept $3200$
states. Averages were calculated over $100$ disorder realizations. We
show in Fig.~\ref{fig:HCBcorrfct}(a,b) the HCB correlation function for
$\rho_{\rm HCB}=0.25$ and several values of the disorder, $W$, both on
(a) semi-log and (b) log-log scales. This shows clearly that for $W=1$
the decay is a power law while for $W=2.5,\,3,\,3.5$ the decay is
exponential. This confirms that the critical disorder in this limit,
$W_c^{\rm HCB}$ is finite. To determine $W_c^{\rm HCB}$, we follow the
same procedure explained above for the fermion case.  We show in
Fig.~\ref{fig:HCBDisorderrho0.25} the collapse of $D_s$ curves
obtained at various systems sizes giving $W_c^{\rm HCB} = 1.5$. Note
that since a hard core boson represents a pair, then the fermionic
system at $\rho=0.5$ should be compared with the HCB system at
$\rho_{\rm HCB}=0.25$. In addition, to do this calculation, we took $2t^2/U = 1$ in
Eq.~\eqref{eq:hardcoreham}. Therefore, to compare this
result with $W_c$ for the spinful system at very large $U$ where the
effective HCB model is accurate, we must multiply $W_C^{\rm HCB}$ by
$2/|U|$. Agreement at large $U$ is excellent, Fig.\ref{fig:WcDs}(b), confirming the accuracy of the effective HCB model and also confirming
that $W_c$ is nonzero. This also confirms that the critical Luttinger
parameter that applies in this $U$ regime is that of hard core bosons
on a crossed chain, and not that of spinful fermions.

In Ref.\cite{Liang2023Disorder}, $W_c$ was calculated at $U=8$ and $\rho = 0.25$ for the
same system we study here and the authors found $W_c(U=8) = 0$ whereas we find
$W_c=0.131$ (Fig.~\ref{fig:CreutzDisorderU8rho0.25}) and that $W_c\neq 0$ for all values of $U$. A possible
reason for this disagreement is the approximation used in Ref.\cite{Liang2023Disorder} to calculate $D_s$, Eq.~\eqref{eq:Ds}.  As mentioned above, in the pure case, the ground state energy is periodic
in the applied phase twist with period $\pi$, $E_\text{GS}(\Phi) =
E_\text{GS}(\Phi+\pi)$, and $E_\text{GS}(0 < \Phi \leq \pi/2)$ is quadratic in
$\Phi$. In this case using the approximation $D_s \approx
\frac{8L}{\pi}E_\text{GS}(\pi/2)-E_\text{GS}(0))$ is justified. However, we show in Fig.~\ref{fig:EG_Phi_Disorder} that in the presence of disorder, $E_\text{GS}(\Phi)$ is no
longer quadratic at large $\Phi$, rendering inavlid this approximation. This possibly underestimates $D_s$, leading to the conclusion that $W_c = 0$.

\subsection{Mean Field Theory Results}\label{sec:MFT}

\begin{figure}
    \centering
    \includegraphics[width=8.6cm]{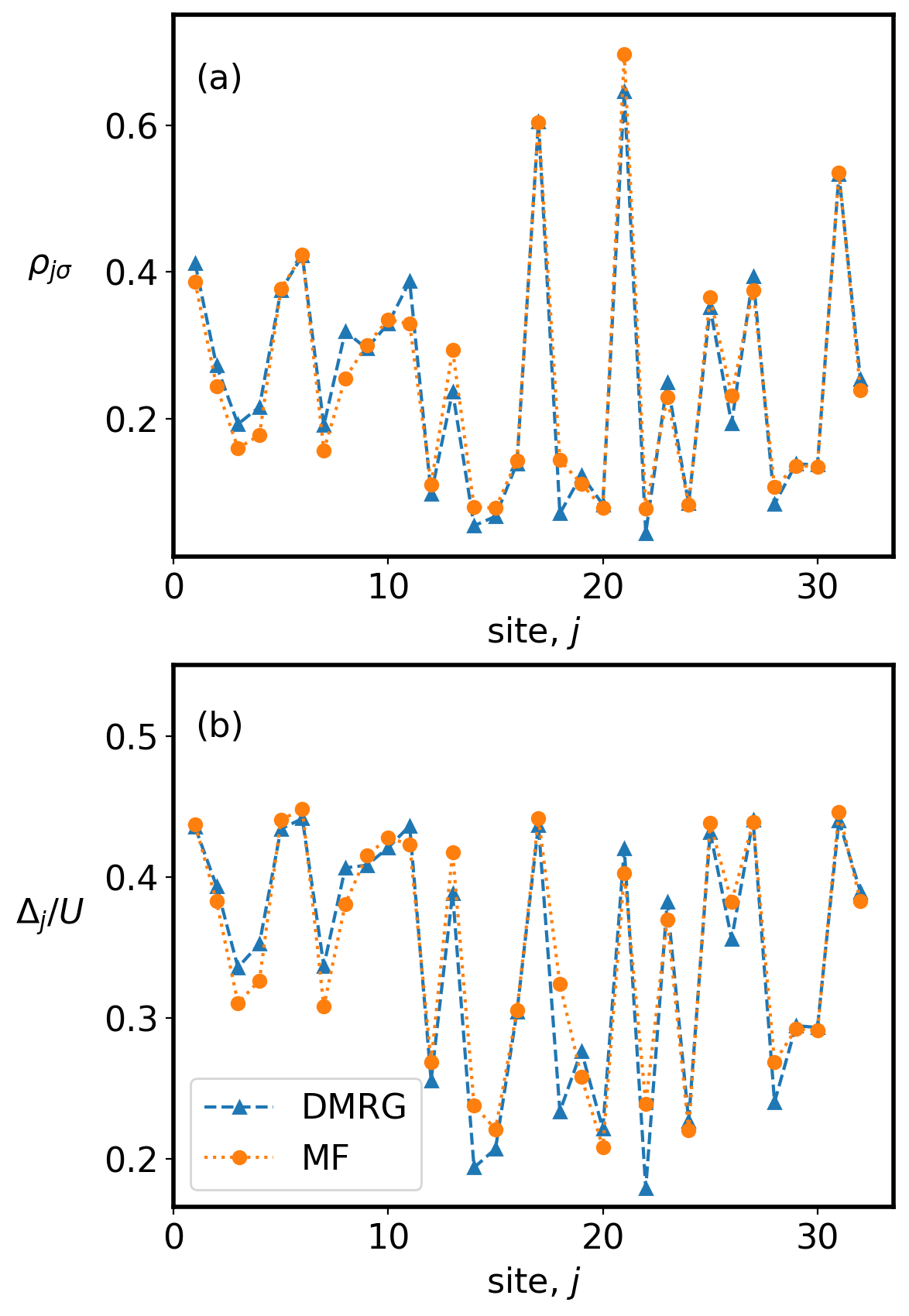}
    \caption{(Color online) The site-dependent density and pairing
      order parameter profiles for a single disorder realization with
      $W=0.6$, at $U=8$, $L = 16$ and $\rho = 0.5$. Site density is
      shown for both sublattices, and dotted lines are provided to
      guide the eye. For $U=8$ and $\rho =0.5$, the critical disorder
      is $W_c=0.272$.}
    \label{fig:DMRGMFdensityprofile}
\end{figure}

In the absence of disorder, multi-band mean field theory (MFT) has
shown remarkable agreement with exact DMRG results in previous studies
of quasi one-dimensional flat band
systems~\cite{chan2022designer,chan2022pairing,tovmasyan2016effective} and quantum Monte Carlo for quasi two-dimensional flat band systems~\cite{julku2016geometric,
  iglovikov2014superconducting,huhtinen2022revisiting}. In the
presence of disorder, MFT becomes more subtle and should be treated
carefully. As we have shown above, the delocalized-localized quantum
phase transition is in the BKT universality class, which MFT cannot
capture. Instead, MFT will predict a second-order phase
transition. Consequently, one should not expect very good quantitative
agreement of $D_s$ between DMRG and MFT in the critical region and in
the localized phase. Nevertheless, it is still worthwhile to study the
results of MFT for this system and what it predicts for $W_c$. In
Appendix~\ref{appendix:MFT} we review briefly the multiband MFT.

\begin{figure}
    \centering
    \includegraphics[width=8.6cm]{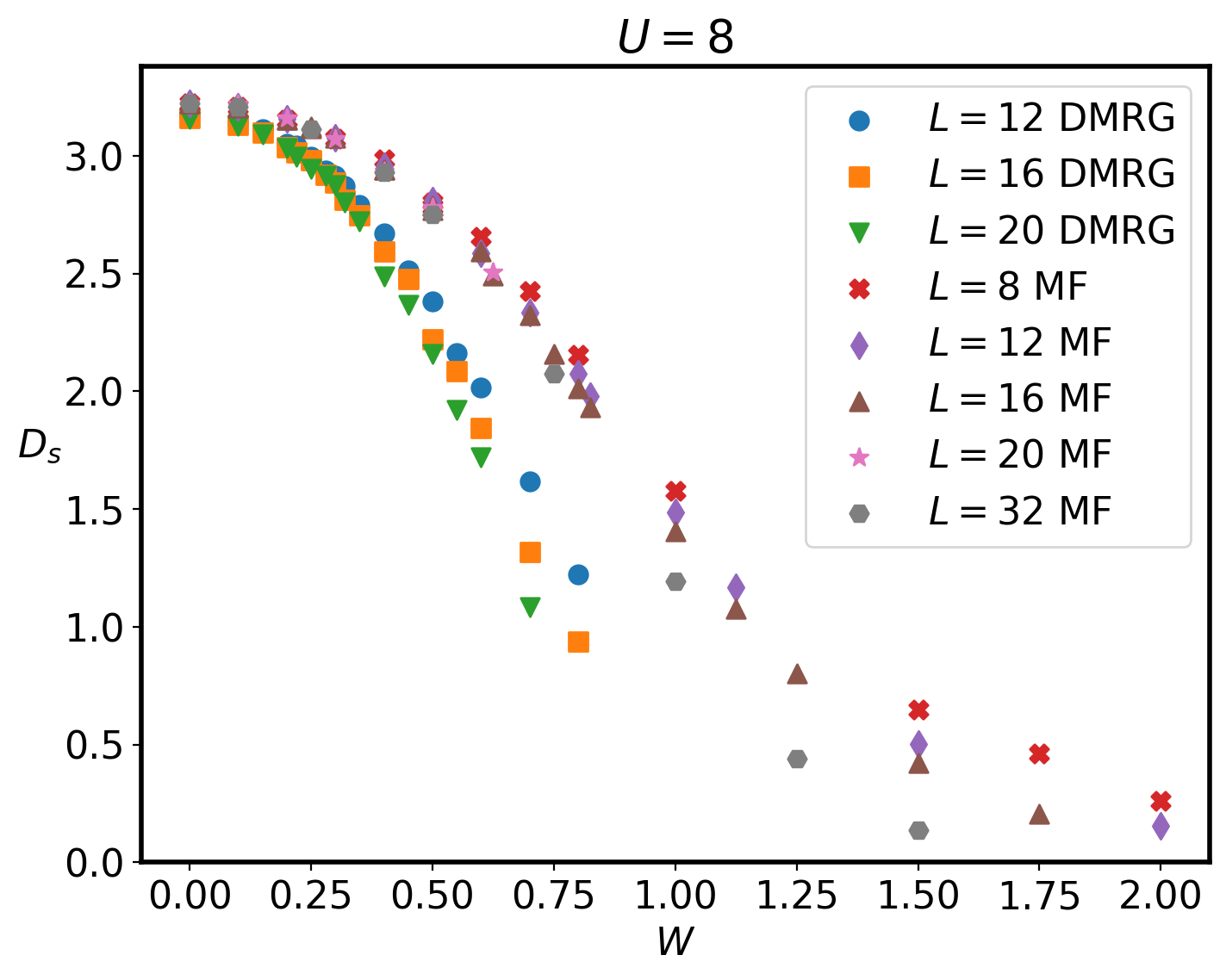}
    \caption{(Color online) $D_s$ as a function of disorder strength
      $W$ showing both DMRG and MF results at $U=8$. The difference in
      $D_s$ is evident quantitatively, especially in the localization
      regime.}
    \label{fig:DMRG_MF_Ds}
\end{figure}

\begin{figure}
    \centering
    \includegraphics[width=8.6cm]{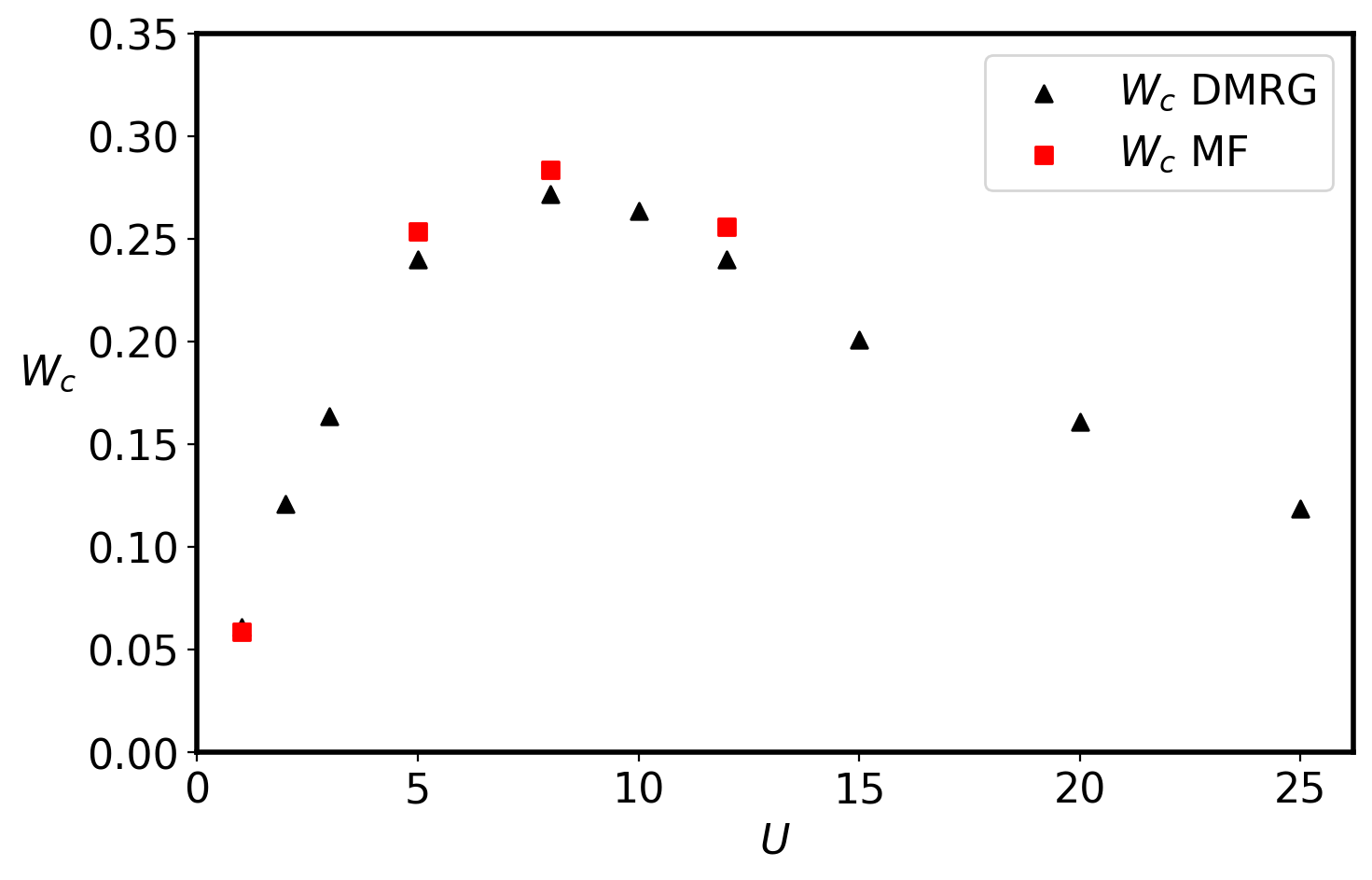}
    \caption{(Color online) $W_c$ as a
      function of $U$ calculated with DMRG and MFT agree well, using appropriate scaling functions.}
    \label{fig:DMRGMF_Wc}
\end{figure}

\begin{figure*}
    \centering
    \includegraphics[width=\textwidth]{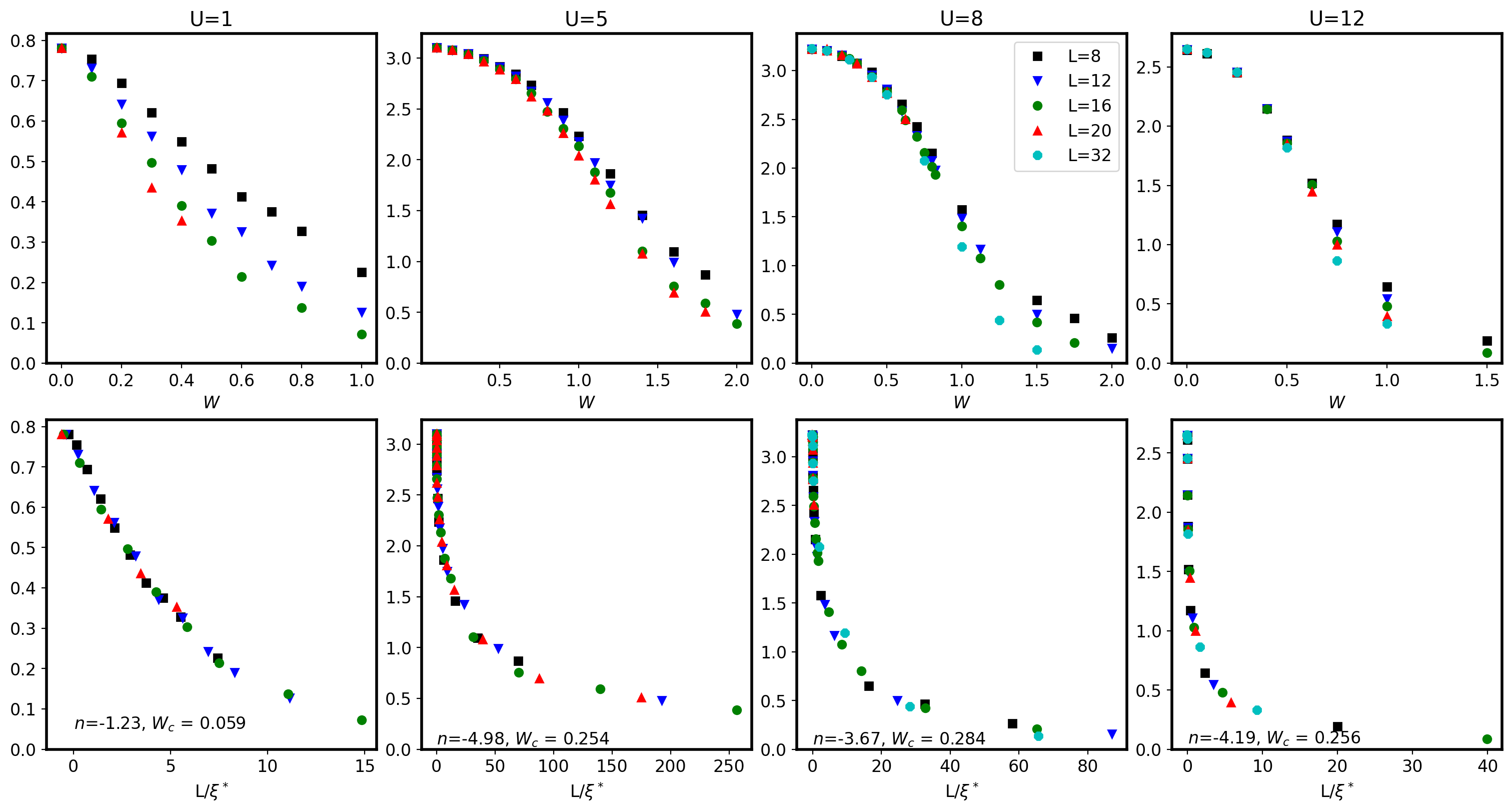}
    \caption{(Color online) \textbf{Top}: $D_s$ versus $W$ and \textbf{Bottom}: $D_s$ versus $L/\xi^*$ where $\xi^* = \abs{W-W_c}^{-\mathit{n}} $ for $U = 1, 5, 8, 12$ and fermion density $\rho=0.5$, calculated using MFT.}
    \label{fig:DsmanyUMF}
\end{figure*}

We first show in Fig.~\ref{fig:DMRGMFdensityprofile}(a) the density
profile, $\rho_{j\sigma} = \langle
c^\dagger_{j\sigma}c^{\phantom\dagger}_{j\sigma}\rangle$,
and (b) the pairing order parameter profile, $\Delta_j/U = \langle
c_{j\uparrow}c_{j\downarrow}\rangle$, for one
realization of disorder. These profiles are for a system with $L=16$
({\it i.e.} a total of $32$ sites), $U=8$, $\rho=0.5$, and $W=0.6$
which is greater than the critical disorder, $W_c=0.272$. These
profiles are in the localized phase and show excellent agreement
between DMRG and MFT. However, disagreement appears in transport
quantities such as the superconducting weight, $D_s$, shown in
Fig.~\ref{fig:DMRG_MF_Ds} as a function of $W$ for several values of
$L$. We see that DMRG and MFT agree up to a value of $W$ close to
$W_c=0.272$ at which point disagreement appears due to the different
universality classes of the transitions, BKT for DMRG and second order for MFT.
This emphasizes that the MFT nevertheless manages to capture the underlying physics of the phase transition, namely the phase coherence of the pairs, and thereby the pair superfluidity, is destroyed by the disorder, even though the (average) pair density remains finite. This is in sharp contrast with the usual MFT behavior, where, the vanishing of the superfluid density is associated with the loss of pairing.

In spite of this disagreement in the values of $D_s$ for $W
\gtrsim W_c$, when the MF curves are scaled appropriately with $\corrdisorder^*=\abs{W-W_c}^{-\mathit{n}}$, to calculate $W_c$, shown in
Fig.~\ref{fig:DsmanyUMF}, the critical disorder calculated using MFT is very close to values obtained with DMRG. (Fig.~\ref{fig:DMRGMF_Wc}). Therefore, even
though MFT predicts the wrong universality class for the delocalized-localized quantum phase transition, it predicts an important quantity $W_c$ accurately.

\section{Conclusions}\label{sec:concs}

In this study, we demonstrated how to compute accurately the disorder
localization length, $\corrdisorder$, and the superfluid weight,
$D_s$, in the presence of site impurities. Additionally, we emphasize
the importance of obtaining the actual $\corrdisorder$, which may
provide more information about the scaling function and the
persistence of phase coherence in disordered \textit{finite size}
systems. By doing so, we confirmed that the delocalized-localized
phase transition scaling function is of the BKT form, with a prefactor
which is vital in quantitative accuracy. On the other hand, if one
were only concerned with finding the critical disorder, $W_c$, and has
determined the scaling form, then minimizing the cost function remains
effective. In addition, good agreement and consistency was
demonstrated between the BKT scaling form for the correlation functions and the data collapse of $D_s$.

We have demonstrated that superconducting transport on the flat band
is, in fact, resilient to on-site disorder, and the critical disorder
is proportional to $D_s$ of the clean system. This provides an
additional benefit to the existing promise of flat band
superconductivity.

We also showed that although MFT cannot exhibit the BKT transition
between the localized and delocalized phases, and predicts instead a
second order transition, the values of $W_c^{\rm MFT}$ are in good
agreement with $W_c$ given by DMRG if the correct power law scaling
function is used.

A natural extension of this work would then be to investigate the
robustness against other types of disorder and its dependence on the
band gap. Since there are different length scales governing the
disordered flat band system, \textit{i.e.} the localization length due
to the band flatness, the correlation length between pairs and the
disorder localization length, one should also determine their relative
magnitudes. In previous work, $D_s$ was found to be significantly
enhanced at low $U$ for flat bands with $E_\text{gap}=0$ to the next
dispersive band \cite{julku2016geometric,chan2022designer}. This,
along with our finding that $W_c \propto D_s$ suggests that flat band
systems with $E_\text{gap}=0$ to the next dispersive band will
similarly have higher $W_c$ at weak attraction. Lastly, for a
different class of flat band systems, the Wannier-Stark flat bands,
the lattice is effectively $2D$, although the flat bands are quasi
$1D$, and superconductivity is supported by multiple flat
bands~\cite{mallick2021wannier, chan2024superconductivity}. It would
be interesting to identify the dependence of flat band
superconductivity on the strength of interactions and disorder in
these systems, and to figure out if there is a critical disorder
strength.

As this work was nearing completion we became aware of reference
\cite{bouzerar2025}. In this paper the authors use the mean field
approximation to study the robustness of superconductivity against
{\it bond} ({\it i.e.} hopping term) disorder in a two-dimensional
Lieb lattice model.

\underbar{\bf Acknowledgments:} The DMRG
computations were performed with the resources of the National
Supercomputing Centre, Singapore (www.nscc.sg).

\appendix

\section{Mean Field Theory with Disorder}\label{appendix:MFT}

As in Refs.\cite{chan2022pairing, chan2022designer, thumin2023flat},
we use mean field theory (MF) to study superconducting transport on a
flat band, in the presence of disorder. In this case, one needs to
solve the mean field equations in real space, since the disorder is
on-site and the system loses translational invariance. The MF
Hamiltonian is the Bogoliubov-de Gennes Hamiltonian derived in detail
in Ref. \cite{chan2022pairing}, explicitly written as
\begin{equation}
    \begin{aligned}
    H_\mathrm{MF}-\mu N&= \displaystyle\sum_{i,j,\alpha,\sigma}\left(
    t_{ij}^{\alpha,\alpha'}c^{\alpha\dagger}_{i,\sigma}c^{\alpha'}_{j,\sigma}
    + h.c.\right) \\
    &-U\displaystyle\sum_{j,\alpha}\left(\rho^\alpha_{j,\uparrow}
    c^{\alpha\dagger}_{j,\downarrow}c^{\alpha}_{j,\downarrow} +
    \rho^\alpha_{j,\downarrow} c^{\alpha\dagger}_{j,\uparrow}
    c^{\alpha}_{j,\uparrow}\right)\\
    &-\displaystyle\sum_{j,\alpha}\left(\Delta^\alpha_j
    c^{\alpha\dagger}_{j,\downarrow}c^{\alpha\dagger}_{j,\uparrow} +
    \Delta^{\alpha*}_jc^{\alpha}_{j,\uparrow}c^{\alpha}_{j,\downarrow}\right)\\
    &+\displaystyle\sum_{j,\alpha,\sigma} (-\mu + \mu_{j,\alpha})
    c^{\alpha\dagger}_{j,\sigma}c^{\alpha}_{j,\sigma}\\
    &+L\displaystyle\sum_{j,\alpha}
    \left(U\rho^\alpha_{j,\uparrow}\rho^\alpha_{j,\downarrow} +
    \abs{\Delta_j^\alpha}^2/U\right).
    \end{aligned}
\end{equation}
where $\rho^\alpha_{j\sigma} = \langle
c^{\alpha\dagger}_{j\sigma}c^{\alpha\phantom\dagger}_{j\sigma}\rangle$
is the density of spin $\sigma$ fermions on site $j$ of sublattice
$\alpha$, and $\Delta^\alpha_j/U = \langle
c^{\alpha}_{j\uparrow}c^{\alpha}_{j\downarrow}\rangle$
is the pairing order parameter, and both quantities are site-dependent
in the presence of disorder. With MF calculations, the chemical
potential $\mu$ is tuned to obtain the desired lattice filling, while
$\mu_{j,\alpha}$ represents the site disorder. By solving the MF
equations self-consistently, \textit{i.e.} diagonalizing the
Hamiltonian, we can obtain the site-dependent profile of
$\rho_{j,\sigma}^\alpha$ and $\Delta_j^\alpha/U$. To calculate the SC
weight, $D_s$, we add a phase twist to the system as shown in detail
in Ref.~\cite{chan2022pairing} and then use Eq.(\ref{eq:Ds}) and
average over many disorder realizations.

\section{Minimization of the Cost Function}\label{appendix:costfunc}

The cost function is defined
with~\cite{Liang2023Disorder,vsuntajs2020ergodicity,vsuntajs2020quantum}
\begin{equation}
    C=\frac{\sum_i\abs{X_{i+1}-X_{i}}}{\mathrm{max}(X_i)-\mathrm{min}(X_i)}-1
\end{equation}
where $C$ is a function of the scaling parameters, \textit{i.e.} $b$
and $W_c$ with the BKT scaling form. $X_i$ is the $i$-th element of a
set containing all $D_s$ values when $D_s$ is sorted as a function of
$L/\corrdisorder^*$.

\begin{figure}
    \centering
    \includegraphics[width=8.6cm]{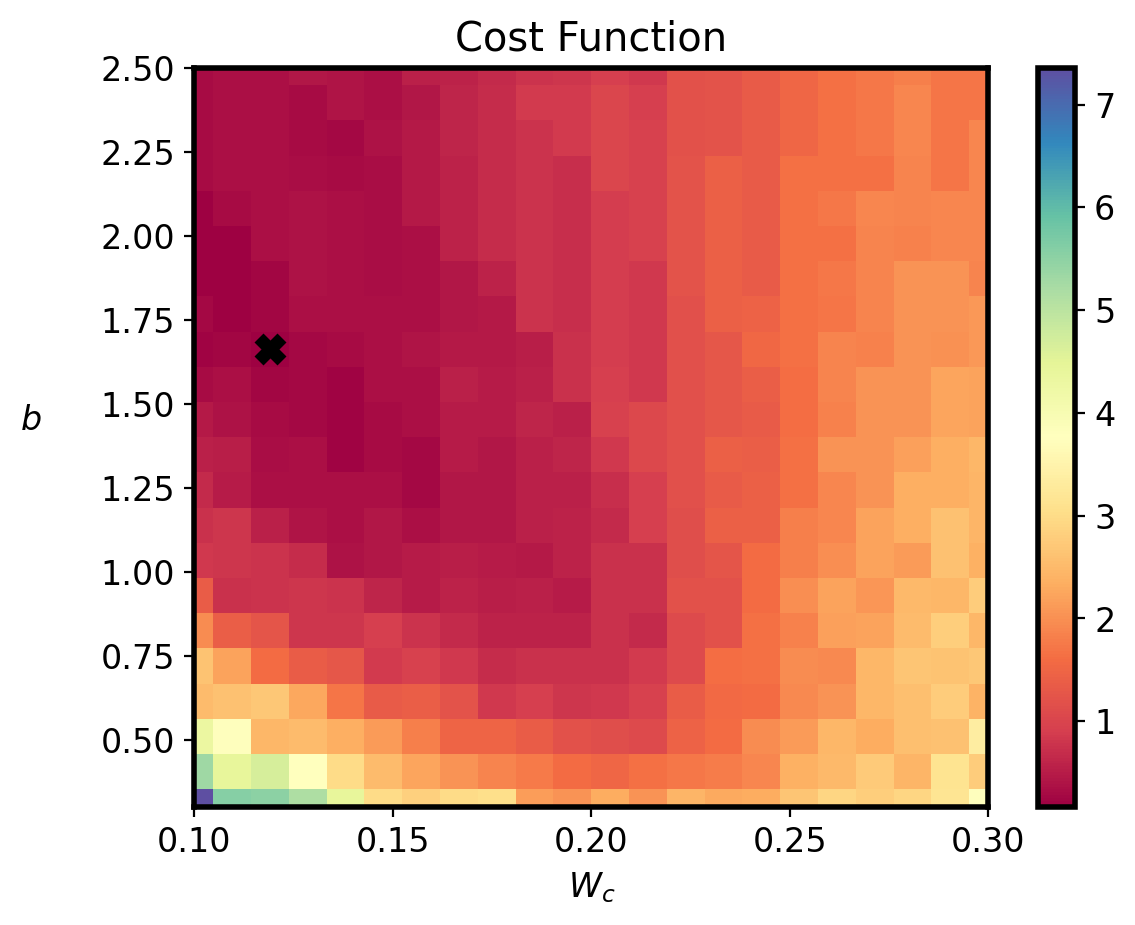}
    \caption{(Color online) With $U=25$ for the Creutz lattice, we
      find that the cost function is minimized when $b=1.66$ and
      $W_c=0.119$.}
    \label{fig:costfunction}
\end{figure}

To find the scaling parameters that optimize the cost function, in
pratice we determine the values of $b$ and $W_c$ that give the minimum
$C$ (Fig.~\ref{fig:costfunction}). Using the values $b=1.66$ and
$W_c=0.119$, we find a good collapse of the $D_s$ values as a function
of $L/\corrdisorder^*$ shown in the main text.

\begin{figure*}
    \centering
    \includegraphics[width=\textwidth]{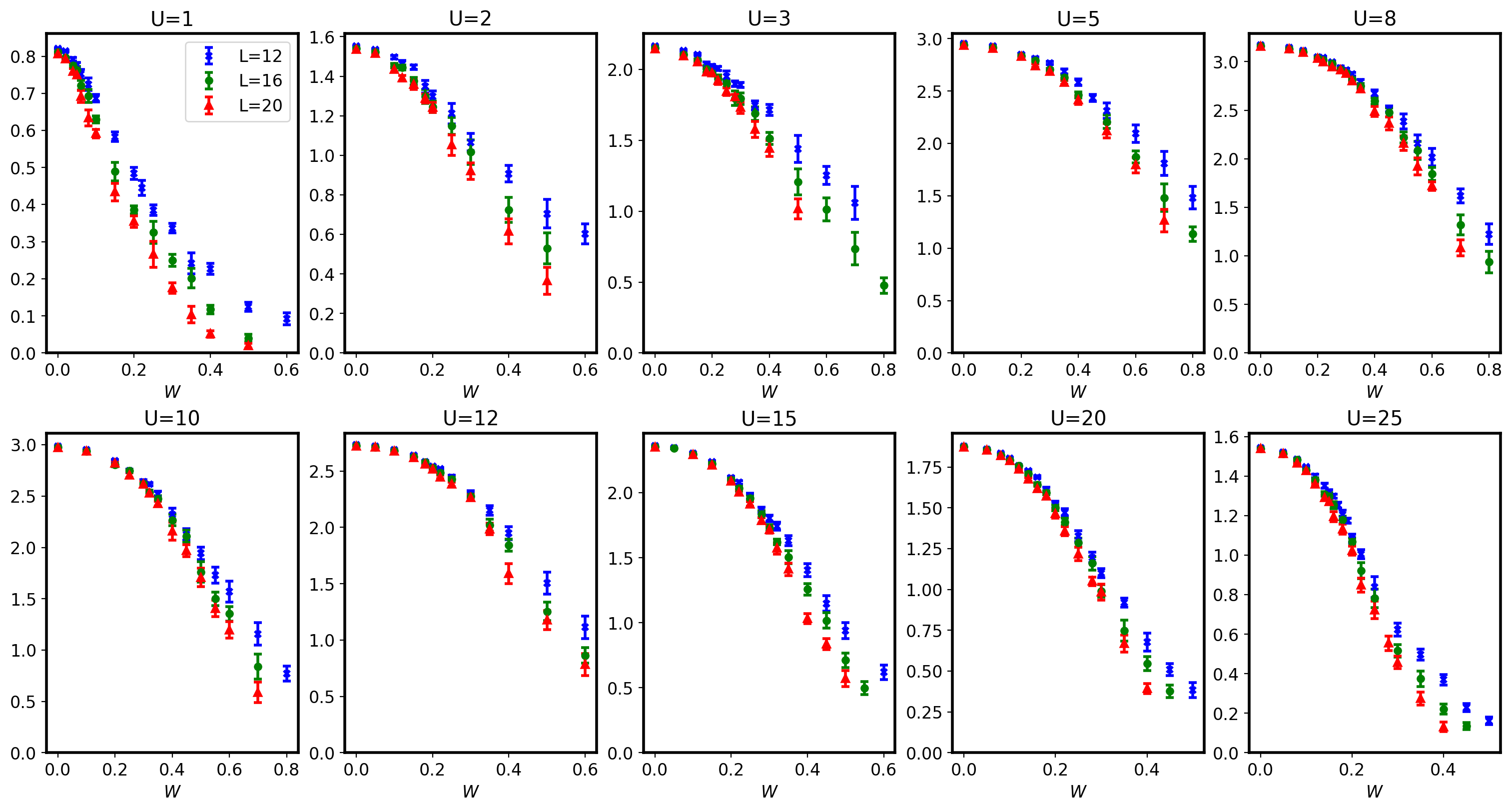}
    \caption{(Color online) $D_s$ versus $W$ for three lattice sizes
      and several values of $U$ for fermion density $\rho=0.5$, calculated using DMRG.}
    \label{fig:DsmanyU}
\end{figure*}

\begin{figure*}
    \centering
    \includegraphics[width=\textwidth]{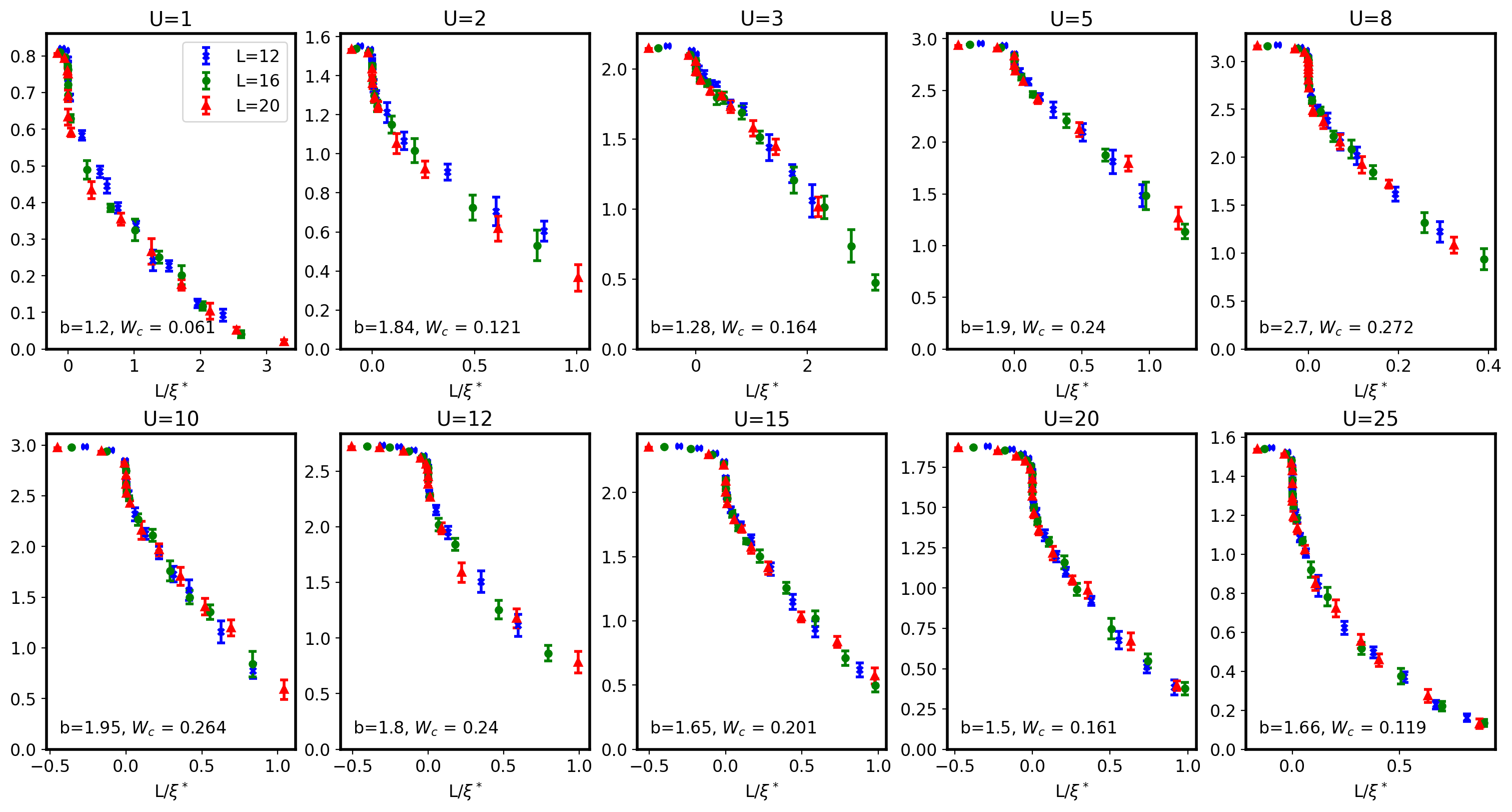}
    \caption{(Color online) $D_s$ as a function of $L/\xi^*$ for three lattice sizes
      and several values of $U$ for fermion density $\rho=0.5$, calculated using DMRG.}
    \label{fig:DsmanyUcollapsed}
\end{figure*}

Finally, we show in Fig.~\ref{fig:DsmanyU} we show $D_s$ versus $W$ for
$L=12,\,16,\,20$ and $1 \geq U \leq 25$ before scaling. In
Fig.~\ref{fig:DsmanyUcollapsed} the same curves are depicted as a
function of the scaling parameter $L/\xi^*$ ($-L/\xi^*$) for $W>W_c$
($W<W_c$) as described above. To reiterate, we found that the BKT
scaling function is consistent between the disorder correlation
length, and the scaling of $D_s$, providing a complete picture of the
disorder induced localization with flat bands.

\clearpage
\bibliography{flatband}

\end{document}